\documentclass[amssymb,amsmath,pra,twocolumn,floatfix,showpacs]{revtex4}
\usepackage{graphicx}

\begin{document}
\title{Resonant and off-resonant transients in electromagnetically induced
transparency: turn-on and turn-off dynamics}
\author{Andrew D. Greentree}
\author{T. B. Smith}
\author{S. R. de Echaniz}
\author{A. V. Durrant}
\affiliation{Quantum Processes Group, Department of Physics and Astronomy, The Open
University, Walton Hall, Milton Keynes, MK7 6AA, UK.}
\author{J. P. Marangos}
\author{D. M. Segal}
\affiliation{Laser Optics and Spectroscopy Group, Blackett Laboratory, Imperial
College of Science Technology and Medicine, Prince Consort Road, London SW7 2BW, UK.}
\author{J. A. Vaccaro}
\affiliation{Department of Physics and Astronomy, University of Hertfordshire,
College Lane, Hatfield, AL10 9AB, UK.}

\begin{abstract}
This paper presents a wide-ranging theoretical and experimental study of
non-adiabatic transient phenomena in a $\Lambda $ EIT system when a strong
coupling field is rapidly switched on or off. The theoretical treatment uses
a Laplace transform approach to solve the time-dependent density matrix
equation. The experiments are carried out in a Rb$^{87}$ MOT. \ The results
show transient probe gain in parameter regions not previously studied, and
provide insight into the transition dynamics between bare and dressed states.
\end{abstract}

\maketitle

\section{\label{Intro}Introduction}

There is now a huge interest in the novel optical properties of
coherently-prepared atomic media, especially electromagnetically induced
transparency (EIT) where a resonant coupling field (or pump field)
coherently prepares an atomic sample to allow dissipation-free propagation
of a weak probe field accompanied by strong dispersion. \ There have been
applications of EIT to enhanced non-linear optical processes \cite{bib:NLO},
laser cooling \cite{bib:EITCool}, quantum non-demolition measurements \cite
{bib:Quant Dem} and gain without inversion\ \cite{bib:LWI}. \ EIT is also
the mechanism underlying the recent experiments in ultra-slow \cite
{bib:SlowLight1} and ultra-fast group velocities \cite{bib:UltraFastLight}
and is potentially of use for storage and retrieval of quantum information
using robust ground state coherences \cite{bib:StoppedLight}. \ There is
also a continuing interest in exploiting the strong optical non-linearities
for the control of light by light at the single photon level \cite{bib:QNLO}%
. \ A review of EIT and its applications can be found in ref \cite
{bib:EITRev}.

The study of transient excitation of three level systems is a mature field,
but there is still a need for experimental verifications of theoretical
work. \ An early theoretical paper by Berman and Salomaa \cite
{bib:Berman1982} compared the dressed-atom and bare-atom pictures, and
considered transients after probe turn-on. \ Related dressed-atom transients
for two level atoms are presented theoretically by Lu and Berman \cite
{bib:Lu1987}. \ Theoretical three-level transient studies considering intial
conditions have also appeared in Lu {\it et al. }\cite{bib:Lu1986}. Harris
and Luo \cite{bib:Harris1995} studied transient EIT in the context of the
energy required for the preparation of EIT. \ Li and Xiao \cite{bib:EITTrans}
investigated the time required for the onset of EIT and\ Zhu considered the
conditions required for observing inversionless gain in the transient regime
for the V \cite{bib:Zhu1996} and $\Lambda $ \cite{bib:Zhu1997} schemes.

Experimental work looking at dressed state transients includes phase
shifting measurements in a two level system \cite{bib:Bai1985}, fluorescence
measurements on a three level $\Lambda $ system \cite{bib:Bai1986} and
pump-probe experiments on a three level ladder system \cite{bib:Lee1987}. \
Transient gain was first observed experimentally in a sodium sample in a $%
\Lambda $ configuration by Fry {\it et al.} \cite{bib:transExp}. \ This is
also the only previous experiment we know of which analyzed transient
dynamics associated with the turn-off of the coupling field. \ RF
experiments on the N-V centre of diamond appear in \cite{bib:Wei}. \ In our
earlier work, we have demonstrated transient EIT \cite{bib:Chen1998} and
Rabi oscillations and gain without inversion \cite{bib:OurTransient} in a
cold rubidium $\Lambda $ system after rapidly switching on a resonant
coupling field.

The work presented here differs from all previous studies in several ways. \
It extends previous work in transient EIT by studying turn-on and turn-off
transients, and\ the region of parameter space investigated is much broader
than previously studied. This is the only work we know of which studies
turn-off transients for off-resonant probe and coupling fields,
inversionless gain away from resonance and transient dressed state
interference away from either bare or dressed state resonance. \ Furthermore
we present new analytical results to describe our experimental transients,
derived using the Laplace transform method, so extending the two level
approach presented in \cite{bib:Schenzle1976}. \ Our theoretical results
also generalize those of Li and Xiao \cite{bib:EITTrans} for mutual
resonance to include two-photon dephasing and arbitrary initial ground state
populations, and we predict a new frequency of oscillation in the probe
transient response. \ We also present an analytical solution, to first order
in the probe strength, for the probe absorption when the coupling field is
non-adiabatically turned off for arbitrary detunings of coupling and probe
fields. \ Our analysis is performed explicitly in the bare state basis. \
Our results provide insight into how atomic systems become dressed by
intense laser fields.

A natural classification of EIT response with switched fields is based on
the relative time scales involved. \ In the {\em adiabatic} regime the
switching is assumed to occur sufficiently slowly that the system evolves
smoothly from one steady state to another. \ For EIT systems where the Rabi
frequency is comparable to the atomic lifetime, the switching is adiabatic
when it occurs on a time scale that is long compared with the relevant
optical pumping times. \ For example, steady state $\Lambda $-type EIT with
resonant fields is characterized by the ground state populations being in
the non-absorbing superposition of the two ground states. \ If the intensity
ratio or relative phase of the two optical fields changes, then the
composition of the non-absorbing superposition state changes, and so the
populations have to be pumped into the new non-absorbing state to maintain
transparency. \ Thus the adiabatic condition requires the switching of the
optical fields to be slow compared with the time taken for optical pumping
from the absorbing state to the non-absorbing state. This is the regime used
in the recent studies of optical information storage and retrieval \cite
{bib:StoppedLight}, the single photon switch \cite{bib:PhotSwitch} and its
classical precursor \cite{bib:YanPreprint}. \ 

There is another adiabatic regime in pulsed systems where evolution occurs
between ground states without spontaneous dissipation. \ In the work by
Harris and Luo \cite{bib:Harris1995} on the conditions for preparing EIT,
this was achieved by using extremely large Rabi frequencies so that all
system evolution to the dark state was achieved on timescales much shorter
than the spontaneous emission time. \ Not requiring such extreme Rabi
frequencies, adiabatic switching using the well-known STIRAP schemes \cite
{bib:STIRAP} decouples the evolution from the spontaneous emission, and so
adiabatic evolution occurs on timescales independent of spontaneous emission
time. The picosecond pulse experiment of Nottleman {\it et al.} \cite
{bib:NottleMann1993} is also effectively in the adiabatic regime. In their
experiment the relative phase of the two Zeeman ground states oscillates by
RF Lamor precession in a magnetic field, and the probe pulses are timed to
arrive at the particular times in the RF cycle when the ground states are in
the non-absorbing state. \ We are not concerned with such regimes in this
work.

In this paper we are concerned with the {\em non-adiabatic} regime, where
the switching time is very fast on the scale of optical pumping times, and
all other relevant time scales, with the coupling field strength comparable
to the spontaneous emission rate. \ The theoretical studies by Li and Xiao 
\cite{bib:EITTrans} and Zhu \cite{bib:Zhu1997} assume (as we do)
instantaneous switching and are therefore in the non-adiabatic regime, as
are the three experimental observations in $\Lambda $ systems \cite
{bib:transExp,bib:Chen1998,bib:OurTransient} where the coupling
field was rapidly switched using a Pockels cell. \ These studies were
restricted to special cases of resonant fields and some specific detunings.

\begin{figure}[tb!]
\includegraphics[bb=82 80 443 708,width=\columnwidth,clip]{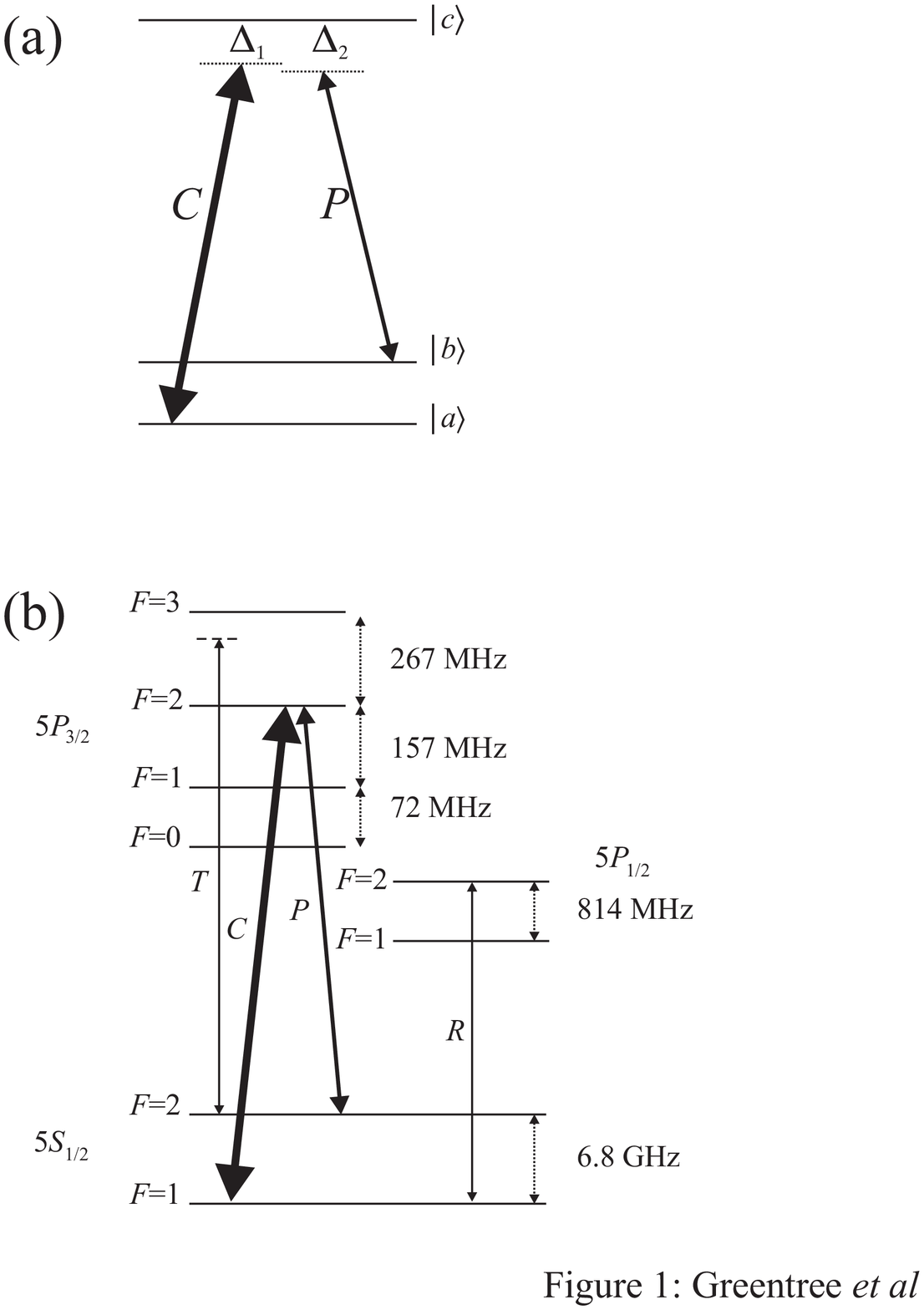}
\caption{\label{Fig1}
(a) Model energy level diagram. \ The atomic energy levels are
labelled $\left| a\right\rangle $, $\left| b\right\rangle $, and $\left|
c\right\rangle $ in order of increasing energy. \ The $\left| a\right\rangle
-\left| c\right\rangle $ transition is driven by a strong coupling field, $C$%
, with detuning, $\Delta _{1}$ and Rabi frequency, $\Omega _{1}\left(
t\right) $. \ The $\left| b\right\rangle -\left| c\right\rangle $ transition
is monitored by a weak probe ,$P$, of detuning, $\Delta _{2}$ and Rabi
frequency $\Omega _{2}<\Gamma $. \ (b) Experimental realization in $^{87}$Rb
MOT. \ The trapping fields $T$ are detuned from the $5S_{1/2}F=1$ to $%
5P_{3/2}F=3$ transition by $\Delta _{T}=-13\mathop{\rm MHz}%
$ and a repumper, $R$, is applied resonantly with the $5S_{1/2}F=1$ to $%
5P_{1/2}F=2$ transition. \ The coupling field is applied to the $5S_{1/2}F=1$
to $5P_{3/2}F=2$ transition and is switched non-adiabatically using a
Pockels cell. The probe is applied to the $5S_{1/2}F=2$ to $5P_{3/2}F=2$
transition. \ Both coupling and probe fields were held at constant detunings
for data collection.}
\end{figure}

There has also been interest in a further non-adiabatic time regime where
effects are studied on timescales comparable with the transition period, see
for example \cite{bib:ShortTime} where the Rabi frequency of the coupling
field was comparable to the transition frequency in a Galium-Arsenide
semiconductor, and in the N-V centre of diamond \cite{bib:RFShortTimes}
where RF transitions at the Rabi frequency were further driven by intense
fields. \ Such regimes are quite different from those considered here as
they go beyond conventional Bloch analysis of the atomic evolution.

The organization of this paper is as follows. \ Section~\ref{Laplace} outlines the
Laplace transform method for calculating transient evolution of an ideal $%
\Lambda $ system, presents some simple analytical results for coupling field
turn-on and turn-off, and gives an overview of the general predictions for
arbitrary detunings of both fields. \ Section~\ref{ExpSetup} describes the experimental
realization of a $\Lambda $ system in laser-cooled rubidium and presents the
experimental details. \ The results of the turn-on and turn-off experiments
are presented in Section~\ref{Results} together with the theoretical predictions. \ The
results are discussed in Section~\ref{Conclusions}.

\section{\label{Laplace}Transients in a $\Lambda $ system using Laplace transforms}

We consider an ideal closed $\Lambda $ system of three non-degenerate levels
excited by monochromatic laser beams, as shown in FIG.~\ref{Fig1}(a). The sample is
assumed to be optically thin and so propagation effects are not considered.
\ The atomic energy levels are labelled in order of increasing energy $%
\left| a\right\rangle $, $\left| b\right\rangle $ and $\left| c\right\rangle 
$. \ The $\left| a\right\rangle -\left| c\right\rangle $ ($\left|
b\right\rangle -\left| c\right\rangle $) transition is coupled by an optical
field with frequency $\omega _{1}$ ($\omega _{2}$) and detuning $\Delta
_{1}=\omega _{1}-\omega _{ca}$ ($\Delta _{2}=\omega _{2}-\omega _{cb}$). \
The transition frequencies are defined $\omega _{\beta \alpha }=\omega
_{\beta }-\omega _{\alpha }$, $\alpha ,\beta =a,b,c$ and in our system $%
\omega _{ba}\ll \omega _{ca},\omega _{cb}$. \ For convenience we also define 
$\Delta _{21}=\Delta _{2}-\Delta _{1}$. \ The state $\left| c\right\rangle $
is an excited state which decays by spontaneous emission to the ground (or
metastable) states $\left| a\right\rangle \ $and $\left| b\right\rangle $\
at rates $\Gamma _{ca}$ and $\Gamma _{cb}$\ respectively.\ There is assumed
to be decay of the (two-photon) coherence on the $\left| a\right\rangle
-\left| b\right\rangle $ transition at a rate $\Gamma _{ba},$ but no
incoherent coupling between these two states. \ The Rabi frequency of field
1 (2) is $\Omega _{1}={\bf d}_{ca}\cdot {\bf E}_{1}$ ($\Omega _{2}={\bf d}%
_{cb}\cdot {\bf E}_{2}$) where ${\bf d}_{\alpha \beta }$ is the
electric-dipole moment of the $\left| \alpha \right\rangle -\left| \beta
\right\rangle $ transition, ${\bf E}_{j}$ is the electric vector of optical
field $j$, and we have chosen units such that $\hbar =1$ so that energies
are measured in units of frequency. \ Optical field 1 is the pump or
coupling field ($C$) and field 2 is the probe ($P$). \ For the purposes of
this work we usually assume that the probe is weak, i.e. $\Omega
_{2}<<\Gamma _{cb}$ although the density matrix equations are presented
without this assumption.

In the bare state basis, the density matrix equations of motion for the
system in the rotating frame are 
\begin{widetext}
\begin{eqnarray}
\dot{\rho}_{aa} &=&\Gamma _{ca}\rho _{cc}+i\frac{\Omega _{1}\left( t\right) 
}{2}\left( \rho _{ca}-\rho _{ac}\right) ,  \nonumber \\
\dot{\rho}_{ab} &=&\left( i\Delta _{21}-\Gamma _{ba}\right) \rho _{ab}+i%
\left[ \frac{\Omega _{1}\left( t\right) }{2}\rho _{cb}-\rho _{ac}\frac{%
\Omega _{2}}{2}\right] ,  \nonumber \\
\dot{\rho}_{ac} &=&\left( -i\Delta _{1}-\frac{\Gamma _{ca}+\Gamma _{cb}}{2}%
\right) \rho _{ac}+i\left[ \left( \rho _{cc}-\rho _{aa}\right) \frac{\Omega
_{1}\left( t\right) }{2}-\rho _{ab}\frac{\Omega _{2}}{2}\right] ,  \nonumber
\\
\dot{\rho}_{bb} &=&\Gamma _{cb}\rho _{cc}+i\frac{\Omega _{2}}{2}\left( \rho
_{cb}-\rho _{bc}\right) ,  \nonumber \\
\dot{\rho}_{bc} &=&\left( -i\Delta _{2}-\frac{\Gamma _{ca}+\Gamma _{cb}}{2}%
\right) \rho _{bc}+i\left[ -\rho _{ba}\frac{\Omega _{1}\left( t\right) }{2}%
+\left( \rho _{cc}-\rho _{bb}\right) \frac{\Omega _{2}}{2}\right] , 
\nonumber \\
\dot{\rho}_{cc} &=&-\left( \Gamma _{ca}+\Gamma _{cb}\right) \rho _{cc}+i%
\left[ \frac{\Omega _{1}\left( t\right) }{2}\left( \rho _{ac}-\rho
_{ca}\right) +\frac{\Omega _{2}}{2}\left( \rho _{bc}-\rho _{cb}\right) %
\right] ,  \nonumber \\
\rho _{\alpha \beta } &=&\rho _{\beta \alpha }^{\ast }  \nonumber \\
1 &=&\rho _{aa}+\rho _{bb}+\rho _{cc}.  \label{eq:DensMat}
\end{eqnarray}
\end{widetext}
The method for switching between bare and dressed bases is given in \cite
{bib:Berman1982}. \ Because we measure probe absorption, which is
proportional to $%
\mathop{\rm Im}%
\left[ \rho _{bc}\left( t\right) \right] $, we concentrate on this component
of the density matrix.

The standard approach to solving Eqs.~(\ref{eq:DensMat}) is to
numerically integrate them (see for example \cite{bib:Zhu1997}). \ This
method has the advantage of conceptual ease and has been used to generate
some of the theoretical turn-on results presented in this work. \ However
this approach has the deficiencies that it is not analytic and therefore
gives minimal insight into the underlying dynamics of the problem, and that
it can suffer cumulative numerical errors. \ Accordingly, we have also
considered a Laplace transform solution which can avoid these difficulties.
\ The Laplace method (in appropriate limits) is more amenable to analysis
and in this work we present what we believe to be several new analytical
results, especially the turn-off equations. \ We find an extra frequency of
oscillation when the system responds to turning on the coupling field, and
suggest the regime where this might be observable. \ Even when the Laplace
method cannot easily produce simple explicit formulae, Eqs.~(\ref
{eq:DensMat}) are linear so that any solution depends only linearly on the
initial conditions, and so avoids the problem of cumulative numerical
errors, making the solutions more robust.

Any solution to Eqs.~(\ref{eq:DensMat}) has the form 
\begin{equation}
\rho _{\alpha \beta }\left( t\right) =\sum_{l}a_{l}e^{b_{l}t}.
\label{eq:GeneralForm}
\end{equation}
Our goal therefore is to determine coefficients $a_{l}$ and $b_{l}$. \ The
difficulty lies in the fact that the coefficients depend on some or all of
the system parameters, including $\Omega _{1}$ and $\Omega _{2}$.

Defining the Laplace transform of $\rho _{\alpha \beta }\left( t\right) $ to
be $r_{\alpha \beta }\left( p\right) $ we have \cite{bib:Laplace} 
\[
r_{\alpha \beta }\left( p\right) =\int_{0}^{\infty }dte^{-pt}\rho _{\alpha
\beta }\left( t\right) 
\]
and the transform of $\dot{\rho}_{\alpha \beta }\left( t\right) $ is $%
pr_{\alpha \beta }\left( p\right) -\rho _{\alpha \beta }^{0}$ where $\rho
_{\alpha \beta }^{0}=\rho _{\alpha \beta }\left( 0\right) $. \ Then
Eqs.~(\ref{eq:DensMat}) generate nine coupled algebraic equations for the 
$r_{\alpha \beta }\left( p\right) $. \ By manipulating these equations, one
finds four closed equations for $r_{ac}$, $r_{ca}$, $r_{bc}$, and $r_{cb}$.
\ One of them is
\begin{widetext}
\begin{multline}
r_{ac}\left( p\right) \left\{ p+i\Delta _{1}+\frac{\Gamma _{ca}+\Gamma _{cb}%
}{2}+\frac{1}{p}\frac{\Omega _{1}^{2}}{4}\left[ 1-s\left( p\right) \right] +%
\frac{\left( \Omega _{2}/2\right) ^{2}}{p-i\Delta _{21}+\Gamma _{ba}}%
\right\}   \\
+r_{ca}\left( p\right) \frac{1}{p}\frac{\Omega _{1}^{2}}{4}\left[ s\left(
p\right) -1\right] -r_{bc}\left( p\right) \frac{1}{p}\frac{\Omega _{1}\Omega
_{2}}{4}s\left( p\right) +r_{cb}\left( p\right) \frac{1}{p}\frac{\Omega
_{1}\Omega _{2}}{4}\left[ s\left( p\right) -\frac{p}{p-i\Delta _{21}+\Gamma
_{ba}}\right]   \\
=\rho _{ac}^{0}-\frac{i}{p}\frac{\Omega _{1}}{2}\left[ \rho _{aa}^{0}+\rho
_{cc}^{0}s\left( p\right) \right] -i\frac{\Omega _{2}}{2}\frac{\rho _{ab}^{0}%
}{p-i\Delta _{21}+\Gamma _{ba}},  \label{eq:LaplaceTransformBitb}
\end{multline}
\end{widetext}
with 
\[
s\left( p\right) =\frac{\Gamma _{ca}-p}{p+\left( \Gamma _{ca}+\Gamma
_{cb}\right) }.
\]
The remaining three equations can be obtained from Eq.~(\ref
{eq:LaplaceTransformBitb}) by (i) taking the complex conjugate of Eq.~(\ref
{eq:LaplaceTransformBitb}), treating $p$ as real, (ii) interchanging all
labels $a$ and $b$, and $1$ and $2$, in Eq.~(\ref
{eq:LaplaceTransformBitb}) and (iii) taking the complex conjugate of the
latter `interchanged' equation. \ Keeping $p$ real in these procedures is
merely a formal device. \ Once the equations are written by the scheme
given, $p$ may take on complex values, as it must when one effects the
Laplace inversion to get $\rho _{\alpha \beta }\left( t\right) $.

The four coupled equations allow for arbitrary initial values $\rho _{\alpha
\beta }^{0}$. \ Finally when $r_{ac}$, $r_{bc}$, $r_{ca}$ and $r_{cb}$ are
known, one has 
\begin{eqnarray}
r_{ab}\left( p\right)  &=&\frac{\rho _{ab}^{0}+i\frac{\Omega _{1}}{2}%
r_{cb}\left( p\right) -i\frac{\Omega _{2}}{2}r_{ac}\left( p\right) }{%
p-i\Delta _{21}+\Gamma _{ba}},  \label{eq:rabfinalsoln} \\
r_{ba}\left( p\right)  &=&\frac{\rho _{ba}^{0}+i\frac{\Omega _{2}}{2}%
r_{ca}\left( p\right) -i\frac{\Omega _{1}}{2}r_{bc}\left( p\right) }{%
p+i\Delta _{21}+\Gamma _{ba}},  \label{eq:rbafinalsoln}
\end{eqnarray}
and the diagonal elements are 
\begin{eqnarray}
r_{aa}\left( p\right)  &=&\frac{1}{p}\left[ \rho _{aa}^{0}+\Gamma
_{ca}r_{cc}\left( p\right) \right]   \nonumber \\
&&+i\frac{1}{p}\frac{\Omega _{1}}{2}\left[ r_{ca}\left( p\right)
-r_{ac}\left( p\right) \right] , \label{eq:raafinalsoln} \\
r_{bb}\left( p\right)  &=&\frac{1}{p}\left[ \rho _{bb}^{0}+\Gamma
_{cb}r_{cc}\left( p\right) \right]   \nonumber \\
&&+i\frac{1}{p}\frac{\Omega _{2}}{2}\left[ r_{cb}\left( p\right)
-r_{bc}\left( p\right) \right] , \label{eq:rbbfinalsoln}
\end{eqnarray}
where 
\begin{eqnarray}
r_{cc}\left( p\right)  &=&\frac{\rho _{cc}^{0}+i\frac{\Omega _{1}}{2}\left[
r_{ac}\left( p\right) -r_{ca}\left( p\right) \right] }{p+\Gamma _{ca}+\Gamma
_{cb}}  \notag \\
&&+\frac{i\frac{\Omega _{2}}{2}\left[ r_{bc}\left( p\right) -r_{cb}\left(
p\right) \right] }{p+\Gamma _{ca}+\Gamma _{cb}}. \label{eq:rccfinalsoln}
\end{eqnarray}
Having solved the equations for the $r_{\alpha \beta }\left( p\right) $, one
generates the $\rho _{\alpha \beta }\left( t\right) $ by Laplace inversion,
namely 
\begin{equation}
\rho _{\alpha \beta }\left( t\right) =\frac{1}{2\pi i}\int_{{\sf C}%
}dpe^{pt}r_{\alpha \beta }\left( p\right) ,  \label{eq:InverseLTforrhoij}
\end{equation}
where the contour ${\sf C}$\ in the complex $p$-plane runs vertically from $%
\sigma -i\infty $ to $\sigma +i\infty $ and $\sigma $ is real and chosen
sufficiently large so that ${\sf C}$ runs to the right of all poles of $%
r_{\alpha \beta }\left( p\right) $.

Equations~(\ref{eq:LaplaceTransformBitb}) to (\ref{eq:rccfinalsoln}) yield
solutions of the form 
\[
r_{\alpha \beta }\left( p\right) =\frac{{\cal C}_{\alpha \beta }\left(
p\right) }{{\cal P}_{\alpha \beta }\left( p\right) },
\]
where the ${\cal C}_{\alpha \beta }\left( p\right) $ are polynomials which
depend on the parameters of the problem, and in particular, linearly upon
one or more of the initial values $\rho _{\alpha \beta }^{0}\left( 0\right) $%
. \ The ${\cal P}_{\alpha \beta }\left( p\right) $ are polynomials too, but
do not depend on the initial values. \ Generally the zeros of ${\cal P}%
_{\alpha \beta }\left( p\right) $ in the complex plane are distinct, so that 
$r_{\alpha \beta }\left( p\right) $ may be expanded as partial fractions 
\begin{equation}
r_{\alpha \beta }\left( p\right) =\sum_{l}\frac{{\cal R}_{\alpha \beta
,l}\left( p\right) }{p-P_{\alpha \beta ,l}},  \label{eq:PartialFractions}
\end{equation}
where $P_{\alpha \beta ,l}$ is the $l^{\text{th}}$ root of the equation $%
{\cal P}_{\alpha \beta }\left( p\right) =0$ and the ${\cal R}_{\alpha \beta
,l}\left( p\right) $ are determined using standard algebra. Using this in
the contour integral, Eq.~(\ref{eq:InverseLTforrhoij}) gives 
\begin{equation}
\rho _{\alpha \beta }\left( t\right) =\sum_{l}{\cal R}_{\alpha \beta
,l}\left( P_{\alpha \beta ,l}\right) e^{P_{\alpha \beta ,l}t}.
\label{eq:rhoijSolnFromPartialFractions}
\end{equation}
Although the principles of this approach are simple, the difficulty lies in
finding the zeros of the ${\cal P}_{\alpha \beta }\left( p\right) $, which
may have to be done numerically or by some approximation. \ Most roots $%
P_{\alpha \beta ,l}$ give rise to damped oscillatory terms or to simple
damping, but should there be a pole at the origin, its contribution does not
decay and represents a stationary state. \ Such a long-time limit, should it
exist, is 
\begin{equation}
\lim_{t\rightarrow \infty }\rho _{\alpha \beta }\left( t\right) ={\cal R}%
_{\alpha \beta }\left( 0\right) =\lim_{p\rightarrow 0}\left[ \frac{p{\cal C}%
_{\alpha \beta }\left( p\right) }{{\cal P}_{\alpha \beta }\left( p\right) }%
\right] .  \label{eq:LongTimeLimit}
\end{equation}
To facilitate matters, at very little cost, we shall henceforth take $\Gamma
_{ca}=\Gamma _{cb}=\Gamma $.

We now present a simple illustrative example of the Laplace transform
method, showing explicit expressions for optical pumping due to the probe
only. This is followed by application of the method to the transients
following turn-on and turn-off of the coupling field. \ The latter solutions
are provided to first order in the probe intensity. \ For simplicity the
turn-on expression is only given for resonant fields, however the turn-off
result is presented for arbitrary detunings. \ This section ends with a
graphical overview and discussion of the general turn-on and turn-off
dynamics.

\subsection{\label{sect:OptPumpProbe}Optical pumping by a probe}

As an illustration, we consider optical pumping by the probe in the absence
of a coupling field ($\Omega _{1}=0$). \ This simple illustrative example
encapsulates the Laplace method. \ A Schr\"{o}dinger equation approach to
this familiar problem appears in \cite{bib:Knight1980}. \ We suppose that
when the probe is turned on, at time zero, all three states may be populated
but that there are no coherences, i.e. $\rho _{\alpha \beta }^{0}$ may be
non-zero for $\alpha =\beta $\ only. \ Then, when Eq.~(\ref
{eq:LaplaceTransformBitb}) and the associated interchanged equations are
solved, one finds that only $\rho _{bc}$, $\rho _{cb}$, $\rho _{aa}$, $\rho
_{bb}$ and $\rho _{cc}(=1-\rho _{aa}-\rho _{bb})$ are in general non-zero. \
Use of Mathematica 4.1 \cite{bib:Mathematica} to do the algebra, yields
\begin{widetext} 
\begin{eqnarray}
r_{bc}\left( p\right)  &=&-\frac{\Omega _{2}}{2}\frac{\left[ \Gamma \left(
1-\rho _{aa}^{0}+\rho _{bb}^{0}\right) +p\left( 2\rho _{bb}^{0}+\rho
_{aa}^{0}-1\right) \right] \left[ i\left( p+\Gamma \right) +\Delta _{2}%
\right] }{D\left( p\right) },  \label{eq:rbc[p]} \\
r_{aa}\left( p\right)  &=&\frac{p\left[ \Gamma +\rho _{aa}^{0}\left( \Gamma
+p\right) -\rho _{bb}^{0}\Gamma \right] \left[ \Delta _{2}^{2}+\left(
p+\Gamma \right) ^{2}\right] +2\left( p+\Gamma \right) \left( \Gamma +2\rho
_{aa}^{0}p\right) \left( \Omega _{2}/2\right) ^{2}}{pD\left( p\right) },
\label{eq:raa[p]} \\
r_{bb}\left( p\right)  &=&\frac{\left[ \Gamma \left( 1-\rho _{aa}^{0}+\rho
_{bb}^{0}\right) +\rho _{bb}^{0}p\right] \left[ \Delta _{2}^{2}+\left(
p+\Gamma \right) ^{2}\right] +2\left( p+\Gamma \right) \left( \rho
_{aa}^{0}-1\right) \left( \Omega _{2}/2\right) ^{2}}{D\left( p\right) },
\label{eq:rbb[p]}
\end{eqnarray}
where 
\[
D\left( p\right) =p\left( p+2\Gamma \right) \left[ \Delta _{2}^{2}+\left(
p+\Gamma \right) ^{2}\right] +2\left( p+\Gamma \right) \left( 2p+\Gamma
\right) \left( \Omega _{2}/2\right) ^{2}.
\]
\end{widetext}
Furthermore, $r_{cb}\left( p\right) $ is obtained by complex conjugation of $%
r_{bc}$, holding $p$ real, and then allowing $p$ to be complex. \ Then the $%
\rho _{\alpha \beta }\left( t\right) $ follow by Eq.~(\ref
{eq:InverseLTforrhoij}). \ Applying the limit in Eq.~(\ref
{eq:LongTimeLimit}) to these quantities shows that all $\rho _{\alpha \beta
}\left( t\right) $ except $\rho _{aa}$ approach zero as $t\rightarrow \infty 
$, and $\rho _{aa}\left( t\right) \rightarrow 1$ in that limit. \ This
simply means that even a weak probe field will pump the system into state $%
\left| a\right\rangle $ after a long time, the usual result. \ A close look,
however, shows that for a weak probe field $\left( \Omega _{2}<\Gamma
\right) $ there are three important time periods. \ Since $D\left( p\right) $
is a fourth degree polynomial in $p$, it has four roots. \ The real parts of
three of them are of order $-\Gamma $ or $-2\Gamma $ but the fourth root is
much smaller. \ To see this, we set $D\left( p\right) =0$ and assume that
terms of order $p^{2}$ and higher are small. \ The fourth root $p_{4}$ is
then approximated by 
\[
p_{4}\approx -\frac{\left( \Omega _{2}/2\right) ^{2}}{\Delta _{2}^{2}+\Gamma
^{2}}\Gamma .
\]
Thus, there is an initial period of damped oscillations in $\rho _{\alpha
\beta }\left( t\right) $, which die out in times of order $\Gamma ^{-1}$ or $%
\left( 2\Gamma \right) ^{-1}$. Optical coherences form in an intermediate
period after the initial dampening, and then a longer period obtains when $%
\rho _{bc}\left( t\right) $, $\rho _{bb}\left( t\right) $ and $\rho
_{cc}\left( t\right) $ very slowly damp to zero and $\rho _{aa}\left(
t\right) $ slowly approaches unity. \ This latter time period derives from
the probe optical pumping. \ For small $\Omega _{2}$, the intermediate
period is long and lies between $1/\left( 2\Gamma \right) $ and $1/p_{4}$. \
For this epoch, Eqs.~(\ref{eq:rbc[p]}),(\ref{eq:raa[p]}) and (\ref{eq:rbb[p]})
are well approximated by expanding them to first order in $\Omega _{2}$. \
When this is done and the limit \ref{eq:LongTimeLimit} is taken, one finds 
\begin{eqnarray*}
\rho _{bc}\left( t\right)  &\rightsquigarrow &-\frac{\Omega _{2}}{4}\frac{1-\rho
_{aa}^{0}+\rho _{bb}^{0}}{\Delta _{2}-i\Gamma }, \\
\rho _{aa}\left( t\right)  &\rightsquigarrow &\frac{1+\rho _{aa}^{0}-\rho _{bb}^{0}}{%
2}, \\
\rho _{bb}\left( t\right)  &\rightsquigarrow &\frac{1-\rho _{aa}^{0}+\rho _{bb}^{0}}{%
2}.
\end{eqnarray*}
More generally, although the equation $D\left( p\right) =0$ is of fourth
order and exactly solvable, numerical values for $\rho _{\alpha \beta
}\left( t\right) $ are perhaps best achieved by numerically finding the
roots $p_{1}$ to $p_{4}$.

\subsection{Turn-on transient with resonant fields}

We now shift the time origin to the instant of turn-on of the coupling field
and use Eqs.~(\ref{eq:LaplaceTransformBitb}) to (\ref{eq:rccfinalsoln}) to
find $%
\mathop{\rm Im}%
\left[ \rho _{bc}(t)\right] $ after $t=0$. \ It is assumed that the probe
has been on for some time in order to establish values for $\rho _{aa}^{0}$, 
$\rho _{bb}^{0}$ and $\rho _{bc}^{0}=\left( \rho _{cb}^{0}\right) ^{\ast }$.
\ From the previous discussion we know that in our case, these will be the
only non-zero values of $\rho _{\alpha \beta }^{0}$. \ The probe is assumed
weak, $\Omega _{2}\ll \Gamma $; therefore we treat the problem to first
order in the probe intensity. \ This corresponds to the experimental
situation to be described in Sections~\ref{ExpSetup} and \ref{Results}. \ For simplicity we here
assume $\Gamma _{ba}$ $=0$ and resonant excitation, i.e. $\Delta _{1}=\Delta
_{2}=0$.\ We then find, \cite{bib:Mathematica}, that the Laplace transform
of $%
\mathop{\rm Im}%
\left[ \rho _{bc}\left( t\right) \right] $ (to first order in $\Omega _{2}$
but all orders in $\Omega _{1}$) follows from 
\begin{widetext}
\begin{equation}
\frac{r_{bc}\left( p\right) -r_{cb}\left( p\right) }{2i}=\frac{%
\mathop{\rm Im}%
\left( \rho _{bc}^{0}\right) p}{p\Gamma +p^{2}+\left( \Omega _{1}/2\right)
^{2}}+\frac{\left( \rho _{aa}^{0}-\rho _{bb}^{0}\right) \Omega _{2}/2}{%
p\Gamma +p^{2}+\left( \Omega _{1}/2\right) ^{2}}-\frac{\Omega _{2}}{2}\frac{%
\Gamma \rho _{aa}^{0}+\left( \rho _{bb}^{0}-1\right) \left( p-\Gamma \right)
+2\rho _{aa}^{0}p}{p\left( p+\Gamma \right) \left( p+2\Gamma \right)
+2\left( 2p+\Gamma \right) \left( \Omega _{1}/2\right) ^{2}}.
\label{eq:LTImrhobc}
\end{equation}
\end{widetext}

It is worth looking at the partial contribution given by the first term on
the right hand side of Eq.~(\ref{eq:LTImrhobc}), namely $%
\mathop{\rm Im}%
\left( \rho _{bc}^{0}\right) p/\left[ p\Gamma +p^{2}+\left( \Omega
_{1}/2\right) ^{2}\right] $ for which the poles are 
\[
p=-\frac{\Gamma }{2}\pm \left( i/2\right) \sqrt{\Omega _{1}^{2}-\Gamma ^{2}}.
\]
\ The inverse transform is easily found to give 
\begin{eqnarray}
\mathop{\rm Im}%
\left[ \rho _{bc}\left( t\right) \right]  &=&%
\mathop{\rm Im}%
\left( \rho _{bc}^{0}\right) e^{-\Gamma t/2}\cos \left( \frac{ft}{2}\right) 
\nonumber \\
&&-%
\mathop{\rm Im}%
\left( \rho _{bc}^{0}\right) e^{-\Gamma t/2}\frac{\Gamma }{f}\sin \left( 
\frac{ft}{2}\right) , \label{eq:Imrhobc:AsLiAndXiao}
\end{eqnarray}
where $f=\sqrt{\Omega _{1}^{2}-\Gamma ^{2}}$. A similar result is given by
Li and Xiao \cite{bib:EITTrans} and corresponds to standard damped Rabi
nutation.

More generally, the contributions of $\rho _{aa}^{0}$ and $\rho _{bb}^{0}$
may not be ignorable, especially if $%
\mathop{\rm Im}%
\left( \rho _{bc}^{0}\right) $ is small or zero. \ Therefore we look at the
full expression, Eq.~(\ref{eq:LTImrhobc}), for which the first two terms
are easily inverted. \ The denominator of the third term, however, is a
cubic polynomial in $p$ and can be somewhat complicated. \ But as we are
primarily interested in the case when the coupling field is strong, namely
when $\Omega _{1}\gg \Gamma $, then the roots, $p_{1}$ and $p_{2}$, in the
first two terms on the right hand side of Eq.~(\ref{eq:LTImrhobc}) can be
estimated as 
\[
p_{1}=p_{2}^{\ast }\approx -\frac{\Gamma }{2}+i\frac{\Omega _{1}}{2},
\]
and those of the third term as 
\begin{eqnarray*}
p_{3} &\approx &-\frac{\Gamma }{2}, \\
p_{4} &=&p_{5}^{\ast }\approx -\frac{5\Gamma }{4}+i\Omega _{1},
\end{eqnarray*}
neglecting terms of order $\Gamma /\Omega _{1}$. \ Note that $p_{1}$ and $%
p_{2}$ give rise to damped oscillations of angular frequency $\Omega _{1}/2$
as seen from Eq.~(\ref{eq:Imrhobc:AsLiAndXiao}), that $p_{3}$ yields a
simple damped term, and that $p_{4}$ and $p_{5}$ give rise to damped
oscillations of angular frequency $\Omega _{1}$. \ Using these roots, it is
a matter of algebra to derive the following approximation to $%
\mathop{\rm Im}%
\left[ \rho _{bc}\left( t\right) \right] $: 
\begin{equation}
\mathop{\rm Im}%
\left[ \rho _{bc}\left( t\right) \right] =\Phi _{12}\left( t\right) +\Phi
_{3}\left( t\right) +\Phi _{45}\left( t\right) ,  \label{eq:ImrhobcasPhis}
\end{equation}
where 
\begin{widetext}
\begin{eqnarray}
\Phi _{12}\left( t\right)  &\approx &e^{-\Gamma t/2}\left\{ 
\mathop{\rm Im}%
\left( \rho _{bc}^{0}\right) \cos \left( \frac{\Omega _{1}t}{2}\right) +%
\frac{\Gamma }{\Omega _{1}}\left[ \left( \rho _{aa}^{0}-\rho
_{bb}^{0}\right) \frac{\Omega _{2}}{\Gamma }-%
\mathop{\rm Im}%
\left( \rho _{bc}^{0}\right) \right] \sin \left( \frac{\Omega _{1}t}{2}%
\right) \right\} ,  \label{eq:Phi12(t)} \\
\Phi _{3}\left( t\right)  &\approx &-\frac{\Omega _{2}}{\Gamma }\frac{%
12\left( 1-\rho _{bb}^{0}\right) }{9+\left( 4\Omega _{1}/\Gamma \right) ^{2}}%
e^{-\Gamma t/2},  \label{eq:Phi3(t)} \\
\Phi _{45}\left( t\right)  &\approx &\frac{\Omega _{2}}{\Gamma }e^{-5\Gamma
t/4}\left[ \frac{12\left( 1-\rho _{bb}^{0}\right) }{9+\left( 4\Omega
_{1}/\Gamma \right) ^{2}}\cos \left( \Omega _{1}t\right) +\frac{1-\rho
_{bb}^{0}-2\rho _{aa}^{0}}{2}\frac{\Gamma }{\Omega _{1}}\sin \left( \Omega
_{1}t\right) \right] .  \label{eq:Phi45(t)}
\end{eqnarray}
\end{widetext}
We have compared \cite{bib:Mathematica} the exact numerical inversion of
Eq.~(\ref{eq:LTImrhobc}) with the approximations in Eqs.~(\ref
{eq:Phi12(t)}) to (\ref{eq:Phi45(t)}). \ The agreement is qualitatively good
even for values of $\Omega _{1}/\Gamma $ greater than $2$.

In Eq.~(\ref{eq:ImrhobcasPhis}), $\Phi _{12}\left( t\right) $ is the
contribution of the poles at $p_{1}$ and $p_{2}$. \ That part of $\Phi
_{12}\left( t\right) $ which is proportional to $%
\mathop{\rm Im}%
\left( \rho _{bc}^{0}\right) $ reduces in the limit of $\Omega _{1}\gg
\Gamma $ to Eq.~(\ref{eq:Imrhobc:AsLiAndXiao}). \ $\Phi _{3}\left(
t\right) $ results from the pole at $p_{3}$ and is strictly monotonically
decreasing with time. \ $\Phi _{45}\left( t\right) $ results from poles $%
p_{4}$ and $p_{5}$ and oscillates (in the limit $\Omega _{1}\gg \Gamma $)
twice as fast as $\Phi _{12}\left( t\right) $. \ In all the transient
experimental and theoretical studies we know (including this present
experimental study), the system behavior is dominated by $\Phi _{12}\left(
t\right) $; however, should $%
\mathop{\rm Im}%
\left( \rho _{bc}^{0}\right) $ be small or zero with approximately equal
ground state populations (i.e. $\rho _{aa}^{0}\approx \rho _{bb}^{0}$), then
the faster oscillations of $\Phi _{45}$ might be observable. \ One way to do
this would be to switch on {\em both} the probe and pump non-adiabatically
at $t=0$. \ In this case we would have (in general) non-zero $\rho _{aa}^{0}$%
, $\rho _{bb}^{0}$ and $\rho _{cc}^{0}=1-\rho _{aa}^{0}-\rho _{bb}^{0}$, but
vanishing off-diagonal components. \ A discussion of the faster oscillations
using the 3D Vector Model \cite{bib:Kasapi,bib:VDRHCH} appears in the Appendix.

Equations~(\ref{eq:Imrhobc:AsLiAndXiao}) to (\ref{eq:Phi45(t)}) were derived
putting $\Gamma _{ba}=0$. \ In experimental realizations, a non-zero value
of $\Gamma _{ba}$ can result from collisions, trap inhomogenieties and other
effects. \ We can generalize Eq.~(\ref{eq:Imrhobc:AsLiAndXiao}) including
this dephasing, to yield 
\begin{eqnarray}
\mathop{\rm Im}%
\left[ \rho _{bc}\left( t\right) \right]  &\approx &-%
\mathop{\rm Im}%
\left( \rho _{bc}^{0}\right) e^{-t\left( \Gamma +\Gamma _{ba}\right) /2}%
\frac{\Gamma -\Gamma _{ba}}{f^{\prime }}\sin \left( \frac{f^{\prime }t}{2}%
\right)  \nonumber \\
&&+%
\mathop{\rm Im}%
\left( \rho _{bc}^{0}\right) e^{-t\left( \Gamma +\Gamma _{ba}\right) /2}\cos
\left( \frac{f^{\prime }t}{2}\right) ,  \label{eq:Imrhobc:AsLiAndXiao:Dephasing} 
\end{eqnarray}
where $f^{\prime }=\sqrt{\Omega _{1}^{2}-\left( \Gamma -\Gamma _{ba}\right)
^{2}}$. \ We can also obtain the general forms of Eqs.~(\ref{eq:Phi12(t)})
to (\ref{eq:Phi45(t)}) (for $\Omega _{1}\gg \Gamma $): 
\begin{widetext}
\begin{eqnarray*}
\Phi _{12}\left( t\right)  &=&e^{-t\left( \Gamma +\Gamma _{ba}\right) /2}
\left[ 
\mathop{\rm Im}%
\left( \rho _{bc}^{0}\right) +\frac{\Omega _{2}\Gamma _{ba}}{3}\frac{5\rho
_{bb}^{0}+\rho _{aa}^{0}-2}{\Omega _{1}^{2}}\right] \cos \left( \frac{\Omega
_{1}t}{2}\right)  \\
&&-\frac{\Omega _{2}}{\Omega _{1}}e^{-t\left( \Gamma +\Gamma _{ba}\right) /2}%
\left[ 
\mathop{\rm Im}%
\left( \rho _{bc}^{0}\right) \frac{\Gamma -\Gamma _{ba}}{\Omega _{2}}+\rho
_{aa}^{0}-\rho _{bb}^{0}\right] \sin \left( \frac{\Omega _{1}t}{2}\right) ,
\\
\Phi _{3}\left( t\right)  &=&-2\frac{\Omega _{2}}{\Omega _{1}^{2}}\left[
\Gamma _{ba}-\frac{3}{8}\left( 1-\rho _{bb}^{0}\right) \left( \Gamma
-4\Gamma _{ba}\right) e^{-\Gamma t/2}\right] , \\
\Phi _{45}\left( t\right)  &=&\frac{\Omega _{2}}{\Omega _{1}}e^{-5\Gamma t/4}%
\left[ \left( \frac{3\Gamma }{4}\frac{1-\rho _{bb}^{0}}{\Omega _{1}}+\frac{%
\Gamma _{ba}}{3}\frac{1-2\rho _{aa}^{0}-\rho _{bb}^{0}}{\Omega _{1}}\right)
\cos \left( \Omega _{1}t\right) +\frac{1-2\rho _{aa}^{0}-\rho _{bb}^{0}}{2}%
\sin \left( \Omega _{1}t\right) \right] .
\end{eqnarray*}
\end{widetext}

\subsection{Turn-off transient for arbitrary detunings of pump and probe}

The turn-off case is relatively simple as only the probe field acts during
the transient period. Thus we are concerned with optical pumping by the
probe, as considered earlier, but in this case the initial values are
different and we shall work to first order in the probe field $\Omega _{2}$.
\ We suppose that both the coupling and probe fields have been turned on for
a time long compared to $\Gamma ^{-1}$ and the coupling field is then
switched off at a time now taken to be $t=0.$ \ We find \cite
{bib:Mathematica} that the inverse Laplace transform of $\rho _{bc}\left(
t\right) $ is
\begin{widetext}
\[
r_{bc}\left( p\right) =\frac{1}{p+\Gamma +i\Delta _{2}}\left[ \rho
_{bc}^{0}+i\frac{\Omega _{2}}{2}\frac{\left( p-\Gamma \right) \left( 1-\rho
_{aa}^{0}\right) -\rho _{bb}^{0}\left( 2p+\Gamma \right) }{p\left( p+2\Gamma
\right) }\right] +O\left( \frac{\Omega _{2}}{\Gamma }\right) ^{3}.
\]
The initial values $\rho _{aa}^{0}$, $\rho _{bb}^{0}$ and $\rho _{bc}^{0}$%
are found by solving for the $r_{\alpha \beta }\left( p\right) $ during the
epoch when both fields act, and then taking the long-time limits, Eq.~(\ref
{eq:LongTimeLimit}). Thus we have
\begin{eqnarray}
\rho _{aa}^{0} &=&O\left( \frac{\Omega _{2}}{\Gamma }\right) ^{2},  \nonumber
\\
\rho _{bb}^{0} &=&1+O\left( \frac{\Omega _{2}}{\Gamma }\right) ^{2}, 
\nonumber \\
\rho _{bc}^{0} &=&\frac{\Omega _{2}}{2}\frac{\Delta _{21}-i\Gamma _{ba}}{%
\left( \frac{\Omega _{1}^{2}}{4}-\Delta _{2}\Delta _{21}+\Gamma _{ba}\Gamma
\right) +i\left( \Delta _{21}\Gamma +\Delta _{2}\Gamma _{ba}\right) }%
+O\left( \frac{\Omega _{2}}{\Gamma }\right) ^{3},  \label{eq:App1:CWLimit}
\end{eqnarray}
so that, 
\[
r_{bc}\left( p\right) =\frac{1}{p+\Gamma +i\Delta _{2}}\left( \rho
_{bc}^{0}-i\frac{\Omega _{2}}{p}\right) +O\left( \frac{\Omega _{2}}{\Gamma }%
\right) ^{3},
\]
This expression contains simple poles at $p=0$ and $p=-\Gamma -i\Delta _{2}$
so that for $t\geq 0$ one has 
\begin{equation}
\rho _{bc}\left( t\right) \approx \rho _{bc}^{0}e^{-\Gamma t}e^{-i\Delta
_{2}t}-i\frac{\Omega _{2}}{2}\left( \frac{1-e^{-\Gamma t}e^{-i\Delta _{2}t}}{%
\Gamma +i\Delta _{2}}\right) .
\label{eq:Rhobc:TimeEvolutionWithInitialConditions}
\end{equation}
Finally, using Eq.~(\ref{eq:App1:CWLimit}) for $\rho _{bc}^{0}$, and
taking the imaginary part of $\rho _{bc}\left( t\right) $, gives to order $%
\Omega _{2}/\Gamma$
\begin{eqnarray}
\mathop{\rm Im}%
\left[ \rho _{bc}\left( t\right) \right]  &=&-\frac{\Omega _{2}}{2}\frac{%
\Gamma +e^{-\Gamma t}\left[ \Delta _{2}\sin \left( \Delta _{2}t\right)
-\Gamma \cos \left( \Delta _{2}t\right) \right] }{\Gamma ^{2}+\Delta _{2}^{2}%
}  \nonumber \\
&&-\frac{\Omega _{2}}{2}e^{-\Gamma t}\frac{\left[ \Gamma \Delta
_{21}^{2}+\Gamma _{ba}\left( \Gamma \Gamma _{ba}+\Omega _{1}^{2}/4\right) %
\right] \cos \left( \Delta _{2}t\right) }{\left( \Gamma \Gamma _{ba}-\Delta
_{2}\Delta _{21}+\Omega _{1}^{2}/4\right) ^{2}+\left( \Delta _{21}\Gamma
+\Delta _{2}\Gamma _{ba}\right) ^{2}}  \nonumber \\
&&-\frac{\Omega _{2}}{2}e^{-\Gamma t}\frac{\Delta _{21}\left[ \frac{\Omega
_{1}^{2}}{4}-\Delta _{2}\left( \Delta _{21}+\Gamma _{ba}^{2}/\Gamma \right) %
\right] \sin \left( \Delta _{2}t\right) }{\left( \Gamma \Gamma _{ba}-\Delta
_{2}\Delta _{21}+\Omega _{1}^{2}/4\right) ^{2}+\left( \Delta _{21}\Gamma
+\Delta _{2}\Gamma _{ba}\right) ^{2}}, \label{eq:App1:Imrhobct}
\end{eqnarray}
\end{widetext}
which is the required turn-off transient. \ It is interesting to note that $%
\Gamma _{ba}$ enters Eqs.~(\ref
{eq:Rhobc:TimeEvolutionWithInitialConditions}) and (\ref{eq:App1:Imrhobct})
only through the initial conditions. \ This may be understood by recalling
that $\Gamma _{ba}$ refers to a two photon dephasing, and so will not be
dynamically important when only the probe field is acting. \ We note that,
at $t=0$ (the instant of turn-off) one has, from Eqs.~(\ref
{eq:App1:CWLimit}) or (\ref{eq:App1:Imrhobct}),
\begin{widetext}
\[
\mathop{\rm Im}%
\left( \rho _{bc}^{0}\right) =-\frac{\Omega _{2}}{2}\frac{\Gamma \Delta
_{21}^{2}+\Gamma _{ba}\left( \Gamma \Gamma _{ba}+\Omega _{1}^{2}/4\right) }{%
\left( \Gamma \Gamma _{ba}-\Delta _{2}\Delta _{21}+\Omega _{1}^{2}/4\right)
^{2}+\left( \Delta _{21}\Gamma +\Delta _{2}\Gamma _{ba}\right) ^{2}},
\]
\end{widetext}
which is the general expression for the steady-state EIT lineshape for
arbitrary detunings.

\subsection{Overview of transient response}

\begin{figure*}[tb!]
\includegraphics[bb=18 149 526 658,clip]{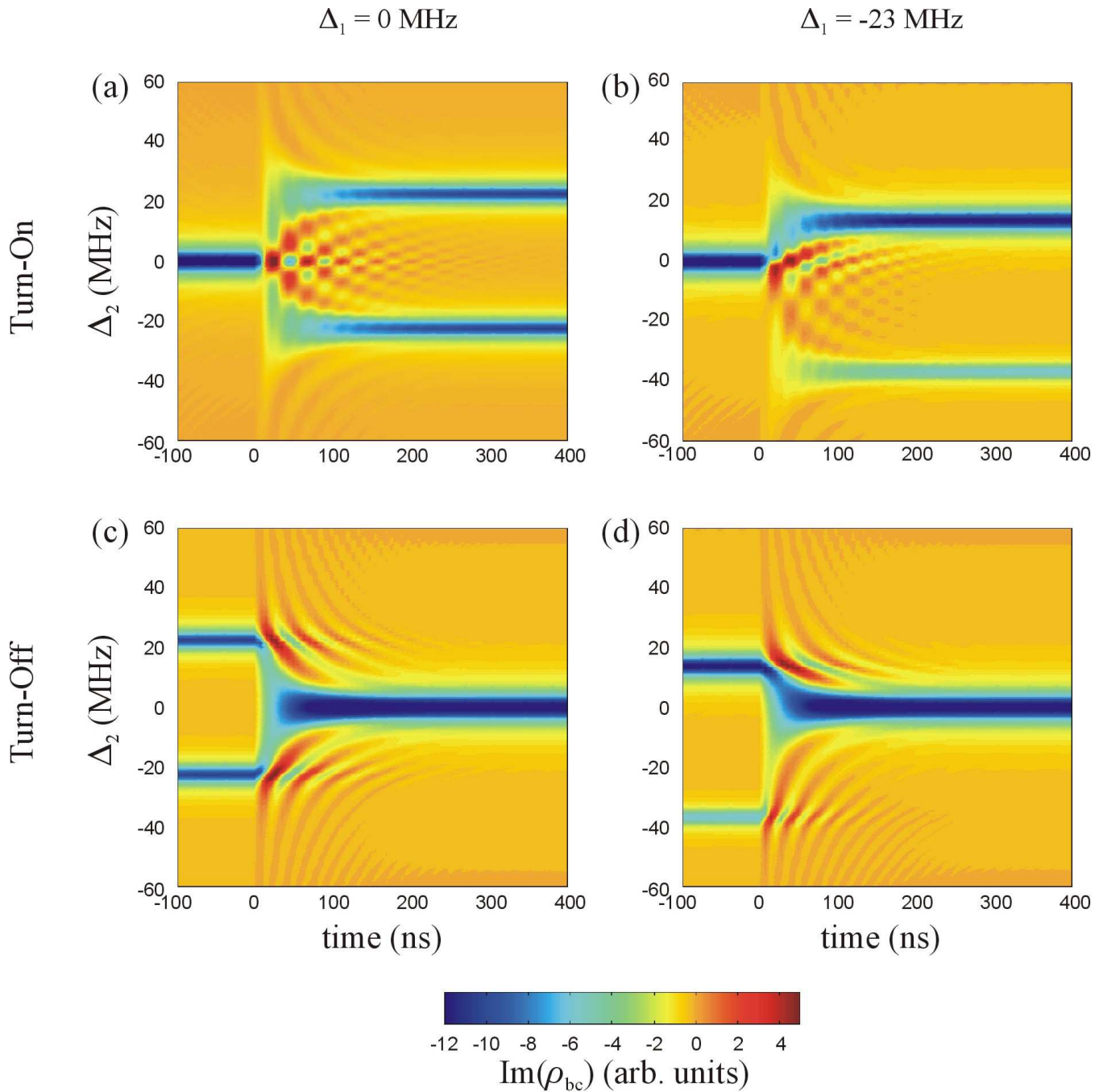}
\caption{\label{Fig2}
Pseudocolour plots of $\mathop{\rm Im}\left[ \rho _{bc}\left( t\right) \right] $%
as a function of time, $t$, and
probe detuning, $\Delta _{2}$ for (a) turn-on; resonant coupling field: $%
\Delta _{1}=0\mathop{\rm MHz}$, (b) turn-on; detuned coupling field: $\Delta _{1}=-23%
\mathop{\rm MHz}$ (c) turn-off; resonant coupling field: $\Delta _{1}=0%
\mathop{\rm MHz}$ and (d) turn-off; detuned coupling field: $\Delta _{1}=-23%
\mathop{\rm MHz}$. \ In all cases we have used: $\Omega _{1}=45%
\mathop{\rm MHz}$, $\Omega _{2}=1\mathop{\rm MHz}$, $\Gamma =5.68%
\mathop{\rm MHz}$ and $\Gamma _{ba}=3.4\mathop{\rm MHz}%
$.\ \ (The effect of uncoupled absorptions are not included in these plots.)}
\end{figure*}

We now present an overview of the turn-on and turn-off transients displayed
as pseudocolour plots showing $%
\mathop{\rm Im}%
\left[ \rho _{bc}\left( t\right) \right] $ as a function of time $t$ and
probe detuning $\Delta _{2}$. \ The parameters are chosen to correspond
closely to the experimental conditions described in detail in the following
sections.  We use the spontaneous radiative decay rate $\left( 2\Gamma =5.68%
\mathop{\rm MHz}%
\right) $ and the effects of uncoupled absorptions are not included in the
model. \ (Uncoupled absorptions are extra absorptions which derive from
degenerate energy levels: see for example \cite{bib:OurFLA} for a full
description). \ In order to generate the pseudocolour plots, we numerically
turned on the probe field at a finite negative time ($t=-10\Gamma $) which
was sufficient to build up a steady state $\rho _{bc}^{0}$. \ Initial
populations (at $t=-10\Gamma $) were chosen so that the populations at time $%
t=0$ corresponded closely to the experimental conditions ($\rho _{aa}\left(
t=-10\Gamma \right) =0.05$, $\rho _{bb}\left( t=-10\Gamma \right) =0.95$). \
We term this the `linear regime' as small variations will only introduce
extra global scaling factors to the amplitude of the probe response.

The turn-on transients are shown in FIGS.~\ref{Fig2}(a) and (b), where the coupling
field is switched on at $t=0.$\ \ We consider first the case of a resonant
coupling field, FIG.~\ref{Fig2}(a).\ \ Scanning the entire figure from left to
right, we first see (for $t<0$) the Lorentzian probe absorption feature of
spectral width $2\Gamma $ as the probe drives the bare $\left|
b\right\rangle -\left| c\right\rangle $ transition. \ Then in the transient
regime from $t=0$ to several $\Gamma ^{-1}$, we find modified Rabi nutations
due primarily to the coupling of the ground state $|b\rangle $ to the
dressed states $|\pm \rangle =\left( |a\rangle \pm |c\rangle \right) /\sqrt{2%
}$. \ If the probe were to couple $|b\rangle $ to a single dressed state,
say $|+\rangle $, it would undergo decaying Rabi oscillations at the detuned
frequency $\sqrt{\left( \Delta _{2}+\Omega _{1}/2\right) ^{2}+\Omega _{2}^{2}%
}$. \ The peaks of these oscillations would lie on the curves $\Delta
_{2}=\pm 2\pi n/t+\Omega _{1}/2$ for $t\ll 2\pi /\Omega _{2}$ and $%
n=1,2,\ldots $. \ Oscillations of this kind are clearly evident in FIG.~\ref
{Fig2}(a) for each of the dressed states. \ However, the fact that the probe
couples $|b\rangle $ to both dressed states simultaneously, evidently,
results in {\em interference }between the Rabi oscillations due to the probe
interacting with each dressed state separately. \ This suggests that the
central pattern in FIG.~\ref{Fig2}(a) represents transient dressed-state
interferences. \ In addition, for $\Delta _{2}\approx 0$, the preparation of
the system before the coupling field is switched on induces relatively
strong absorption and a larger value of $%
\mathop{\rm Im}%
(\rho _{bc})$ at $t=0$. \ This leads to enhanced Rabi nutations along the $%
\Delta _{2}=0$ line for $t>0$.  Finally, for long times, we see the two
well-resolved absorption features, each of width $\Gamma $, corresponding to
the well known Autler-Townes doublet \cite{bib:A-T}. \ We note that our pseudocolour
plots show a superficial similarity to the three dimensional plots
presented by Lu {\it et al }\cite{bib:Lu1986} but the different
initial conditions between the two cases studied result in quite different
dynamics.

We now focus on the case where the probe too is fixed on resonance, i.e. the
time-line $\Delta _{2}=0$ in FIG.~\ref{Fig2}(a).\ \ Here, for $t>0$, we see gain
and absorption cycles of the kind discussed in refs \cite{bib:EITTrans,%
bib:Zhu1997} and observed by us previously \cite{bib:Chen1998,%
bib:OurTransient}.\ For a strong coupling field ($\Omega _{1}/\Gamma \gg 1$%
) this transient ($\Delta _{2}=0$)\ is well described by Eq.~(\ref
{eq:Imrhobc:AsLiAndXiao:Dephasing}). \ Two other lines of note are at $\Delta
_{2}=\pm \Omega _{1}/2$ where transparency for $t<0$ changes transiently to
absorption as the probe begins to resonantly monitor a transition to a
dressed state.

Figure~\ref{Fig2}(b) shows the case of a detuned coupling field. \ Again, we observe
a transient regime featuring absorption and gain regions evolving to the
resolved dressed state absorption features centered at $\Delta _{2}=\Delta
_{1}/2\pm \sqrt{(\Omega _{1}^{2}+\Delta _{1}^{2})/4}$, but in this case we
can clearly distinguish between the dominant or major dressed state and the
narrower minor dressed state. \ In the bare state picture, in the limit of
large $\Delta _{1}$ these two features are associated, respectively, with
one-photon absorption ($\left| b\right\rangle -\left| c\right\rangle $) and
two-photon absorption ($\left| b\right\rangle -\left| c\right\rangle -\left|
a\right\rangle $).

We now consider the turn-off transients shown in FIGS.~\ref{Fig2}(c) and (d), where
both fields are assumed to have been on for all negative times to establish
the initial conditions, and the coupling field is switched off at $t=0.$ \
The dominant feature in the transient regime is the $1/t$ broadened
absorption region where the probe begins to drive the bare transition. Of
more interest are the strong transient gain peaks that can be seen along the
two lines $\Delta _{2}=\Delta _{1}/2\pm \sqrt{(\Omega _{1}^{2}+\Delta
_{1}^{2})/4})$ at the levels associated with the two dressed state
transitions before turn-off. \ This is simply the off-resonant probe
monitoring the relaxation of a two-level atom with specially prepared
initial conditions. \ We note that these transient gain regions are tunable
by varying the detuning and strength of the coupling beam, and are remote
from any uncoupled absorptions in a real $\Lambda $ system that might exist
in the $\Delta _{2}=0$ region \cite{bib:OurFLA,bib:UCA}. \ We note
also that the entire ($t,\Delta _{2}$) region of this figure is described by
Eq.~(\ref{eq:App1:Imrhobct}).

\section{\label{ExpSetup}Experimental set-up and procedure}

\begin{figure}[tbp!]
\includegraphics[bb=45 422 572 569,width=\columnwidth,clip]{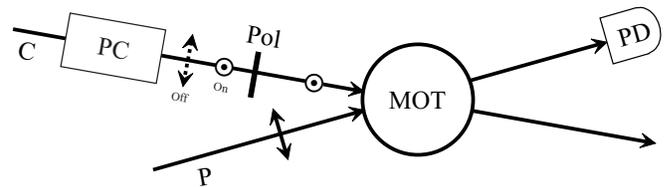}
\caption{\label{Fig3}
Schematic of experimental beam configuration. \ The coupling, $C$,
and probe, $P$, beams make an angle of about $20{{}^\circ}%
$. \ PD is an avalanche photodiode, PC is a Pockels Cell and Pol a
polariser. \ $P$ is linearly polarised in the horizontal plane while $C$ is
linearly polarised in the vertical plane when the PC is on and linearly
polarised in the horizontal plane when the PC is off.}
\end{figure}

The $\Lambda $ system was realized in a Rb$^{87}$ MOT substantially similar
to the one described in \cite{bib:OurTransient}. \ The sample contained
about $5\times 10^{7}$ atoms in a region of diameter about $3%
\mathop{\rm mm}%
$. \ The relevant energy levels and excitation scheme are shown in FIG.~\ref
{Fig1}(b). \ A schematic showing the layout of coupling and probe beams is shown
in FIG.~\ref{Fig3}. \ The coupling beam had an average intensity $\approx 100%
\mathop{\rm mW}%
/%
\mathop{\rm cm}%
^{2}$ with a roughly elliptical profile $2%
\mathop{\rm mm}%
\times 4%
\mathop{\rm mm}%
$. The probe beam had an average intensity $\approx 0.03%
\mathop{\rm mW}%
/%
\mathop{\rm cm}%
^{2}$ with circular profile of diameter about $1%
\mathop{\rm mm}%
$. The beams propagated in the sample with orthogonal linear polarizations
and with an angle of about $20^{0}$ between their directions of propagation.
\ All laser beams were derived from external-cavity laser diodes, using
master-slave injection where appropriate. \ Non-adiabatic switching of the
coupling field was realized using a Linos Photonics LM10 Pockels cell with a
custom designed high voltage switch. \ The rise time of the switch and
associated detection circuitry was less than $6%
\mathop{\rm ns}%
$. \ The excited $5P_{3/2}$ state lifetime is $28%
\mathop{\rm ns}%
$ and the relevant Rabi periods are of similar magnitude and so the
non-adiabatic condition is fairly well satisfied. \ Because the fall time of
the switch is very long, the Pockels cell could only be operated in either
fast turn-on or fast turn-off mode, necessitating separate experiments to
observe turn-on and turn-off transients. \ Changing between turn-on and
turn-off experiments was realized by rotating the plane of polarization of
the coupling field by $90%
{{}^\circ}%
$ before entering the Pockels cell.

The probe detuning was set by varying the cavity size of an external-cavity
grating-controlled diode laser (ECDL) using the voltage offset of the
piezo-electric transducer, which was monitored on a voltmeter. \ This
voltage was calibrated using the saturated absorption spectrum in a standard
Rb cell. \ Drifts in the cavity size due to temperature fluctuations were
checked and corrected for by determining the voltage which put the probe
field on resonance at the beginning of each transient experiment.
Nevertheless, the absolute detuning of the probe field could not be
determined to an accuracy better than about half a linewidth ($\approx 3%
\mathop{\rm MHz}%
$), except when the field was locked to resonance. \ Standard detunings used
were generally much larger than $3%
\mathop{\rm MHz}%
$, with dressed state detunings being as large as $36%
\mathop{\rm MHz}%
$. The measured transients were stored in a Tektronix TDS520B\ digital
oscilloscope. \ Instabilities in our control method limited the maximum
number of averages to about $50$. \ Because of small drifts in the
intensities of our laser diodes (which were run in constant current mode for
maximum frequency and mode stability), probe transmission levels were
measured with the MOT turned off between transient runs and the transmission
level at the time of the experiment inferred using a linear regression.

Our experiments were carried out in the optically thin regime with maximum
probe absorption of about $15\%$ at resonance. \ The probe Rabi frequency of
about $4%
\mathop{\rm MHz}%
$ gave a good signal to noise ratio without producing any measurable power
broadening. \ Under these conditions\ the probe absorption increases
linearly with probe intensity and it was found convenient to normalize our
probe absorption/transmission levels to the steady state absorption in the
absence of the coupling field. \ We use the same method as that described in 
\cite{bib:OurTransient} where we define the maximum probe absorption in the
absence of a coupling field as the zero transmission level. \ Thus, if we
define the voltage measured on our photodetector as ${\cal V}$, the signal
for maximum probe absorption (minimum probe transmission) ${\cal V}_{\min }$
and the signal without any probe absorption ${\cal V}_{0}$, then our scaled
transmission levels for a detected signal are defined by 
\[
T=\frac{{\cal V-V}_{\min }}{{\cal V}_{0}-{\cal V}_{\min }}.
\]
This definition for the transmission has the advantage of being simple and
robust against day-to-day fluctuations in the number of atoms trapped in our
sample (provided we are in a linear absorption regime). \ It is also easy to
make direct comparisons with theory by noting that any detected changes in
signal level will be proportional to $%
\mathop{\rm Im}%
\left( \rho _{bc}\right) $, with values of $T>1$ associated with gain and $%
T<1$ indicating absorption.

With on-resonant or near-resonant probe fields our experimental transients
are accompanied by the time-independent effects of uncoupled absorptions. \
These have been fully described in our previous paper \cite{bib:OurFLA}. \
For the purposes of this work it suffices to treat the system as an ensemble
of identical three level atoms in an ideal $\Lambda $ configuration, as in
Section~\ref{Intro}, together with a similar ensemble that does not interact with the
coupling field. In our turn-on experiments, uncoupled absorptions were found
to contribute about $20\%$ of the maximum resonant absorption and are
assumed to have a Lorentzian line shape centered at zero probe detuning.

\section{\label{Results}Experimental results and discussions}

\begin{figure}[tbp!]
\includegraphics[bb=138 191 458 698,width=\columnwidth,clip]{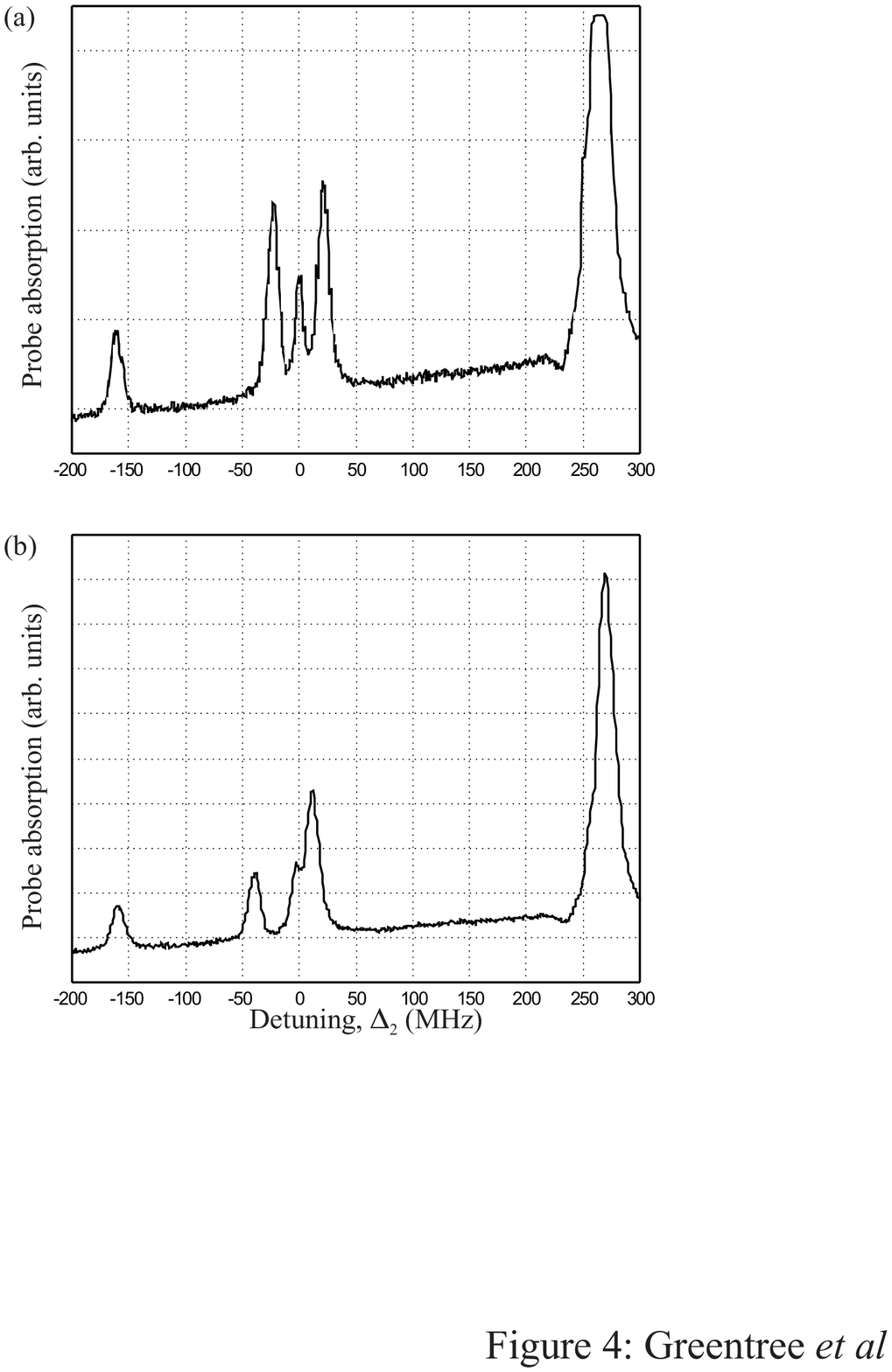}
\caption{\label{Fig4}
Steady state probe transmission spectra. \ (a) On-resonance
coupling field with Rabi frequency $\Omega _{1}=45%
\mathop{\rm MHz}$ showing dressed state absorptions at detunings $\Delta _{2}=\pm \Omega
_{1}/2$ and uncoupled absorption at $\Delta _{2}=0$. \ (The peak on the $%
5S_{1/2},F=2$ to $5P_{3/2},F=3$ transition has been truncated to fit on the
page). \ (b) Off-resonant coupling excitation with Rabi frequency $\Omega
_{1}=45\mathop{\rm MHz}$ and detuning $\Delta _{1}=-23\mathop{\rm MHz}%
$. \ The major dressed state is at $\Delta _{2}=14\mathop{\rm MHz}%
$ and the minor one at $\Delta _{2}=-37\mathop{\rm MHz}%
$. \ Uncoupled absorptions are barely resolved in the wing of the major
dressed state. \ In each case note the presence of extra absorption peaks
corresponding to the $5S_{1/2},F=2$ to $5P_{3/2},F=1$ and $5S_{1/2},F=2$ to $%
5P_{3/2},F=3$ transitions at $\Delta _{2}=-157\mathop{\rm MHz}$ and $\Delta _{2}=267%
\mathop{\rm MHz}$ respectively for frequency calibration purposes. \ Each trace is an
average over $50$ scans. \ Day to day fluctuations in signal levels mean
that the absorption scales in (a) and (b) are different.}
\end{figure}

We first present experimental steady-state probe frequency spectra (EIT
traces) for a resonant and an off-resonant coupling field with the same
parameters that apply in the turn-on transient experiments. \ The central
region of FIG.~\ref{Fig4}(a) shows the probe spectrum for the resonant coupling
field. The dressed state absorption peaks are clearly resolved on either
side of the uncoupled absorption at $\Delta _{2}=0$. \ The Rabi frequency of
the coupling field can be inferred from this figure to be $\Omega _{1}=45%
\mathop{\rm MHz}%
.$ \ This value is consistent with that obtained from beam parameters and
absorption line strengths. \ Also seen are the absorption peaks on the $%
5S_{1/2}F=2$ to $5P_{3/2}F=1$ transition at $\Delta _{2}=-157%
\mathop{\rm MHz}%
$ and the $5S_{1/2}F=2$ to $5P_{3/2}F=3$ transition at $\Delta _{2}=267%
\mathop{\rm MHz}%
.$ \ These peaks were used for frequency calibration purposes. \ Figure~\ref{Fig4}(b)%
{\em \ }shows the probe spectrum when the coupling field is detuned by $%
\Delta _{1}=-23%
\mathop{\rm MHz}%
$. \ The major dressed state absorption peak is seen at $\Delta _{2}=14%
\mathop{\rm MHz}%
$ and the minor one at $\Delta _{2}=-37%
\mathop{\rm MHz}%
$. \ Uncoupled absorptions are barely resolved in the wing of the major
dressed state. \ We note that the linewidths in these spectra are
approximately $11%
\mathop{\rm MHz}%
$, or nearly twice the spontaneous decay rate of the $5P_{3/2}$ levels. \ We
attribute this additional broadening to inhomogeneities in the sample
principally due to light shifts caused by the coupling beam and the trapping
beams with smaller contributions from laser linewidths. \ Accordingly we
have taken a phenomenological value for the half linewidth of $\Gamma =5.5%
\mathop{\rm MHz}%
$ in our theoretical transients rather than the standard value of $2.8%
\mathop{\rm MHz}%
$, as discussed in \cite{bib:OurFLA}. \ The two-photon dephasing rate is
taken to be $\Gamma _{ba}=0.6\Gamma $ \cite{bib:Chen1998}.

We now present experimental turn-on and turn-off transients for the resonant
and the non-resonant coupling field, together with the corresponding
theoretical curves. \ Turn-on theory transients were obtained by numerically
solving Eqs.~(\ref{eq:DensMat}). \ Turn-off theory transients were
obtained by fitting the experimental data to Eq.~(\ref{eq:App1:Imrhobct})%
. \ In each case representative selections of probe detunings are chosen so
that horizontal lines across FIGS.~\ref{Fig2}(a)-(d) are explored.

\begin{figure*}[tbp!]
\includegraphics[bb=92 239 467 683,clip]{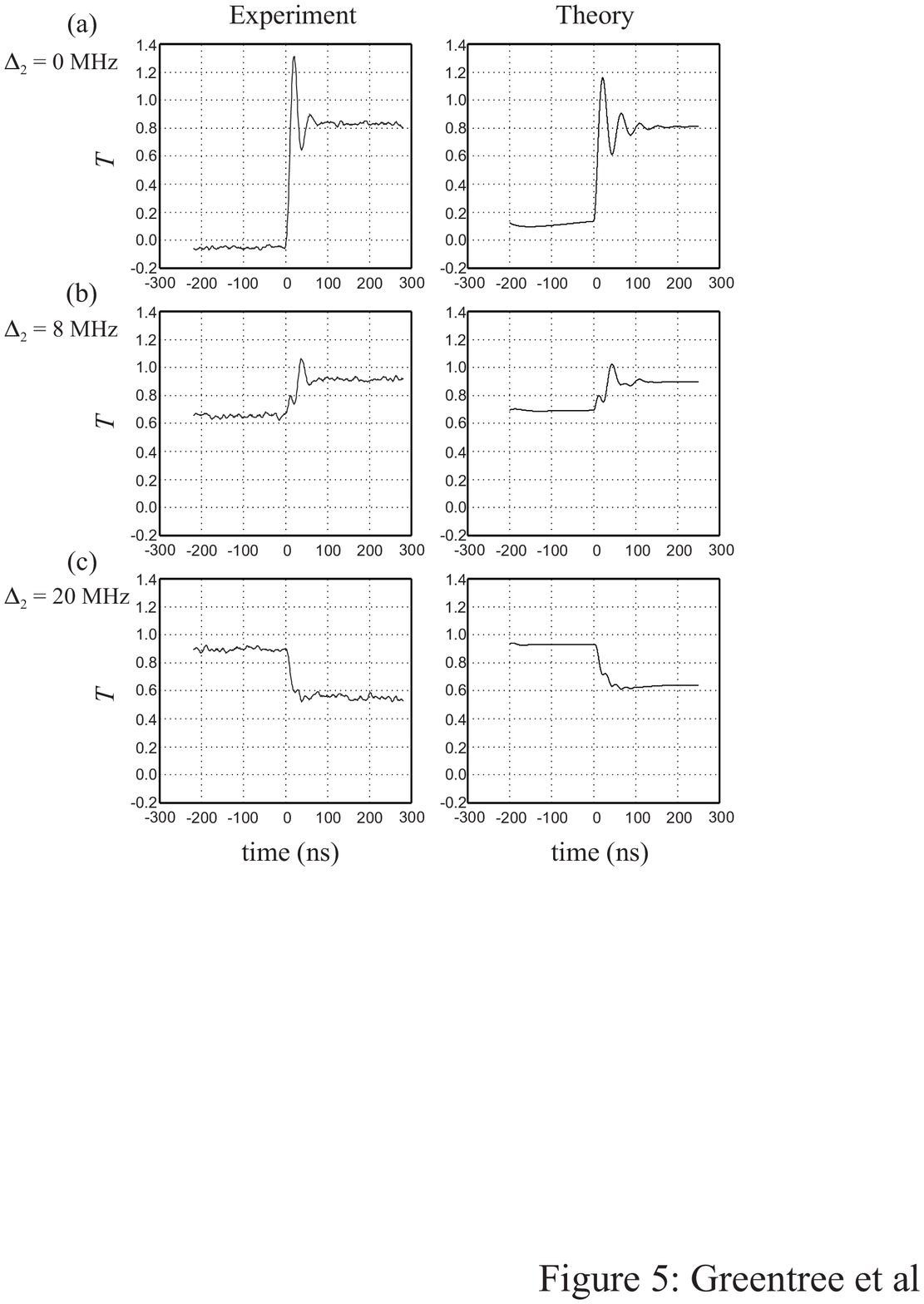}
\caption{\label{Fig5}
Turn-on transients for varying probe detunings with a resonant
coupling field:\ (a) $\Delta _{2}=0\mathop{\rm MHz}%
$ (resonant probe), (b) $\Delta _{2}=8\mathop{\rm MHz}%
$, and (c) $\Delta _{2}=20\mathop{\rm MHz}%
$ (dressed state filling). In all cases $\Omega _{1}=45%
\mathop{\rm MHz}$ and $50$ averages were taken.}
\end{figure*}

When we compare the experimental and theoretical curves in FIGS.~\ref{Fig5}-\ref{Fig8}, we
find excellent agreement in the transient behavior. \ We therefore use our
theoretical results to determine $\Delta _{2}$ which we find to be
consistent with detunings derived from the piezo-offset. \ It has been less
easy to make direct comparisons of the overall transmission levels and there
are minor inconsistencies when comparing experimental with theoretical
levels in our transients. \ We believe that these inconsistencies are due to
difficulties in correcting for all of the intensity fluctuations and drifts
in our lasers on the timescales of our experiments, especially as they
relate to mode hops in the probe laser as it was detuned.

\subsection{Turning on the coupling field}

For these experiments the coupling field is turned on at time zero after the
probe field and MOT fields have been on for a long time. The left-hand
traces in FIGS.~\ref{Fig5}(a)-(c) show the experimental results for turning on a
resonant coupling field ($\Delta _{1}=0$) of Rabi frequency $\Omega _{1}=45%
\mathop{\rm MHz}%
$. \ The right hand traces show the corresponding theory curves, with the
steady transmission levels adjusted to take account of the uncoupled
absorption.\ \ Figure~\ref{Fig5}(a) shows the well-known case of zero probe detuning (%
$\Delta _{2}=0$) exhibiting significant gain ($T=1.35$) despite the presence
of uncoupled absorption. \ This curve is substantially similar to FIG.~3(b)
of \cite{bib:OurTransient}. \ The long-time transmission settles down
to the value of $T=0.8$ showing the effects of uncoupled absorption as well
as effects due to the phenomenological two-photon dephasing rate. \ In
FIG.~\ref{Fig5}(c) the probe detuning is $20%
\mathop{\rm MHz}%
$ and so the probe transmission is almost unity before turn-on. \ After
turn-on, the transmission falls transiently as the probe begins to
resonantly couple the dressed state. \ The long-time transmission is close
to the value $0.5$ which is expected from the equal absorption strengths of
the two dressed states. \ We term this phenomenon, dressed state {\em filling%
}, because the increase in probe absorption and the associated movement of
coherences and populations are accompanying the transient dressing of the $%
\left| a\right\rangle -\left| c\right\rangle $ transition. \ This has
previously been called three-level free-induction decay \cite{bib:Berman1982}%
. \ Figure~\ref{Fig5}(b) shows an intermediate case with probe detuning of $\Delta
_{2}=8%
\mathop{\rm MHz}%
$. \ No significant differences were found between negative and positive
detunings of the probe, as expected from the theory, and so negative
detuning traces are not shown.

\begin{figure*}[tbp!]
\includegraphics[bb=85 54 465 768,scale=.9,clip]{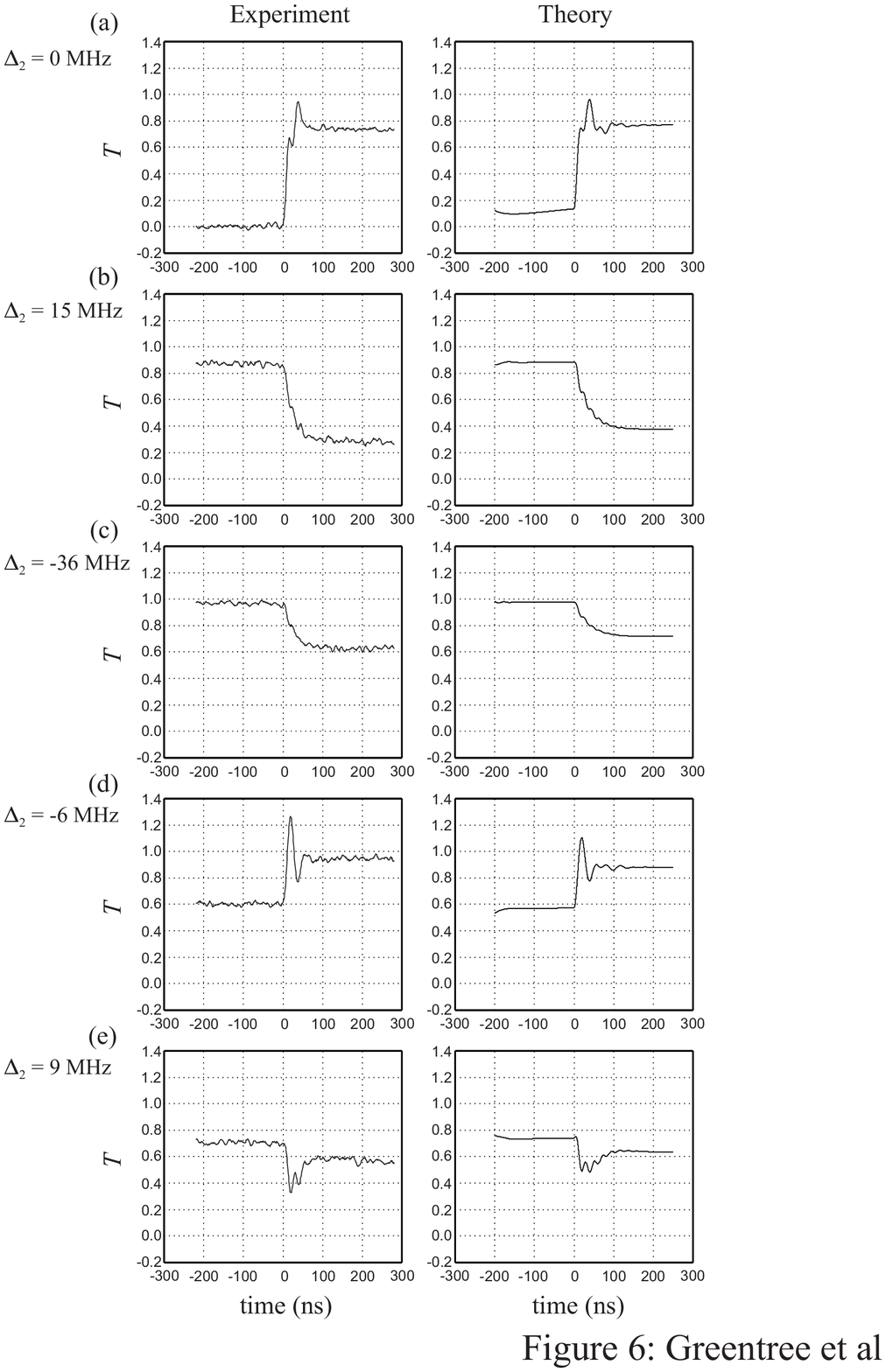}
\caption{\label{Fig6}
Turn-on transients for varying probe detunings with an
off-resonant coupling field:\ (a) $\Delta _{2}=0\mathop{\rm MHz}%
$ (resonant probe), (b) $\Delta _{2}=15\mathop{\rm MHz}%
$ (major dressed state filling), (c) $\Delta _{2}=-36\mathop{\rm MHz}%
$ (minor dressed state filling) (d) $\Delta _{2}=-6\mathop{\rm MHz}%
$ and (e) $\Delta _{2}=9\mathop{\rm MHz}$. \ In all cases $\Omega _{1}=45%
\mathop{\rm MHz}$, $\Delta _{1}=-23\mathop{\rm MHz}$ and $50$ averages were taken.}
\end{figure*}

We now consider the case where the coupling field is detuned from resonance
by $\Delta _{1}=-23%
\mathop{\rm MHz}%
$. \ Other parameters and the dynamics before turn-on are the same as
before. {\em \ }Experimental and theoretical transient results are presented
in FIGS.~\ref{Fig6}(a) to (e). \ The first three figures show (a) the case of a
resonant probe, (b) the filling of the major dressed state ($\Delta _{2}=15%
\mathop{\rm MHz}%
$), and (c) the filling of the minor dressed state ($\Delta _{2}=-37%
\mathop{\rm MHz}%
$). \ Trace (d) ($\Delta _{2}=-6%
\mathop{\rm MHz}%
$) shows the case where the probe is detuned just to the red of the bare
state absorption line. \ It is interesting because of the clarity of the
signal obtained coupled with the relatively high gain ($T\approx 1.26$)
making this a promising region for further transient gain experiments. \
Finally in FIG.~\ref{Fig6}(e) ($\Delta _{2}=9%
\mathop{\rm MHz}%
$) the probe is detuned by a similar amount to the blue and shows enhanced
absorption dips.

\subsection{Turning off the coupling field}

The experimental conditions for turn-off were different in some respects
from those for turn-on. The maximum coupling-field intensity passing the
Pockels cell for turn-off was $\Omega _{1}=46%
\mathop{\rm MHz}%
$, slightly greater than that for turn-on; also the uncoupled absorption
levels are slightly different, probably as a result of optical pumping
amongst the Zeeman substates by the coupling field before turn-off. We have
found that some of the experimental turn-off transients are remarkable for
the high gain and clarity of signal obtained. \ We attribute the clearer
signals to the fact that the system is simpler without the inhomogeneous
light shift effects of the coupling field.

Using the analytical result, Eq.~(\ref{eq:App1:Imrhobct}) with the
addition of the effects of uncoupled absorption, we have been able to
perform fits to the experimental data using Origin 6.0284 \cite{bib:Origin}.
\ The results of these fits are presented as the theoretical traces. \ The
interdependence of many of the parameters mean that some parameters cannot
be adequately determined using this method. \ Fortunately the probe
detuning, $\Delta _{2}$, can be well determined using this method, and the
agreement between experiment and theory shows that our system is well
described by our three level analysis. \ Except where mentioned, the values
for $\Delta _{2}$ quoted below were determined using the curve fitting.

\begin{figure*}[tbp!]
\includegraphics[bb=85 54 465 632,clip]{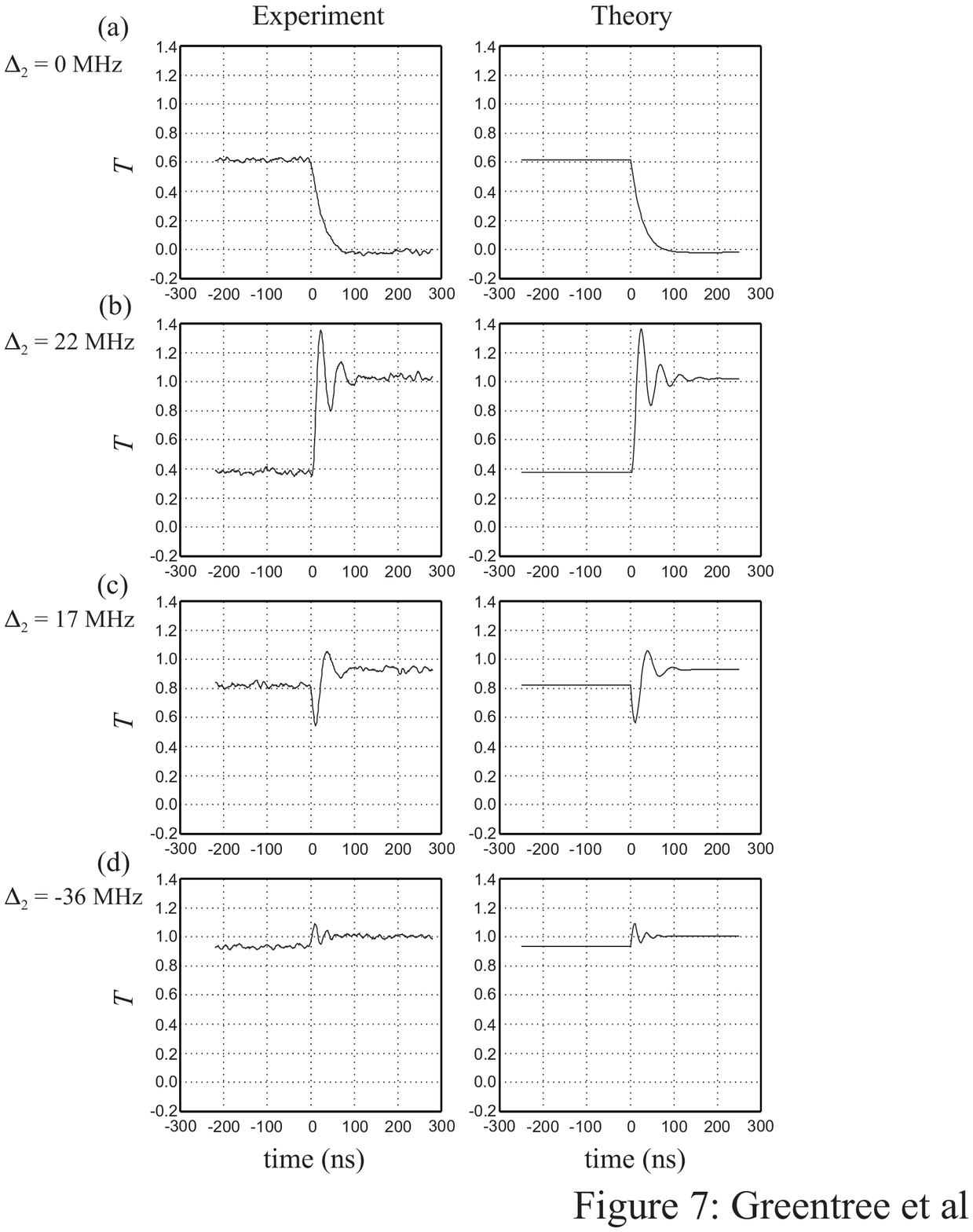}
\caption{\label{Fig7}
Turn-off transients for a strong, on-resonant coupling field: (a) $%
\Delta _{2}=0\mathop{\rm MHz}$ (bare state filling), (b) $\Delta _{2}=22%
\mathop{\rm MHz}$ (dressed state emptying), (c) $\Delta _{2}=17\mathop{\rm MHz}%
$ and (d) $\Delta _{2}=-36\mathop{\rm MHz}$. \ In all cases $\Omega _{1}=46%
\mathop{\rm MHz}$ before turn-off and the uncoupled absorption was found to be
about $20\%$.}
\end{figure*}

\begin{figure*}[tbp!]
\includegraphics[bb=85 54 465 632,clip]{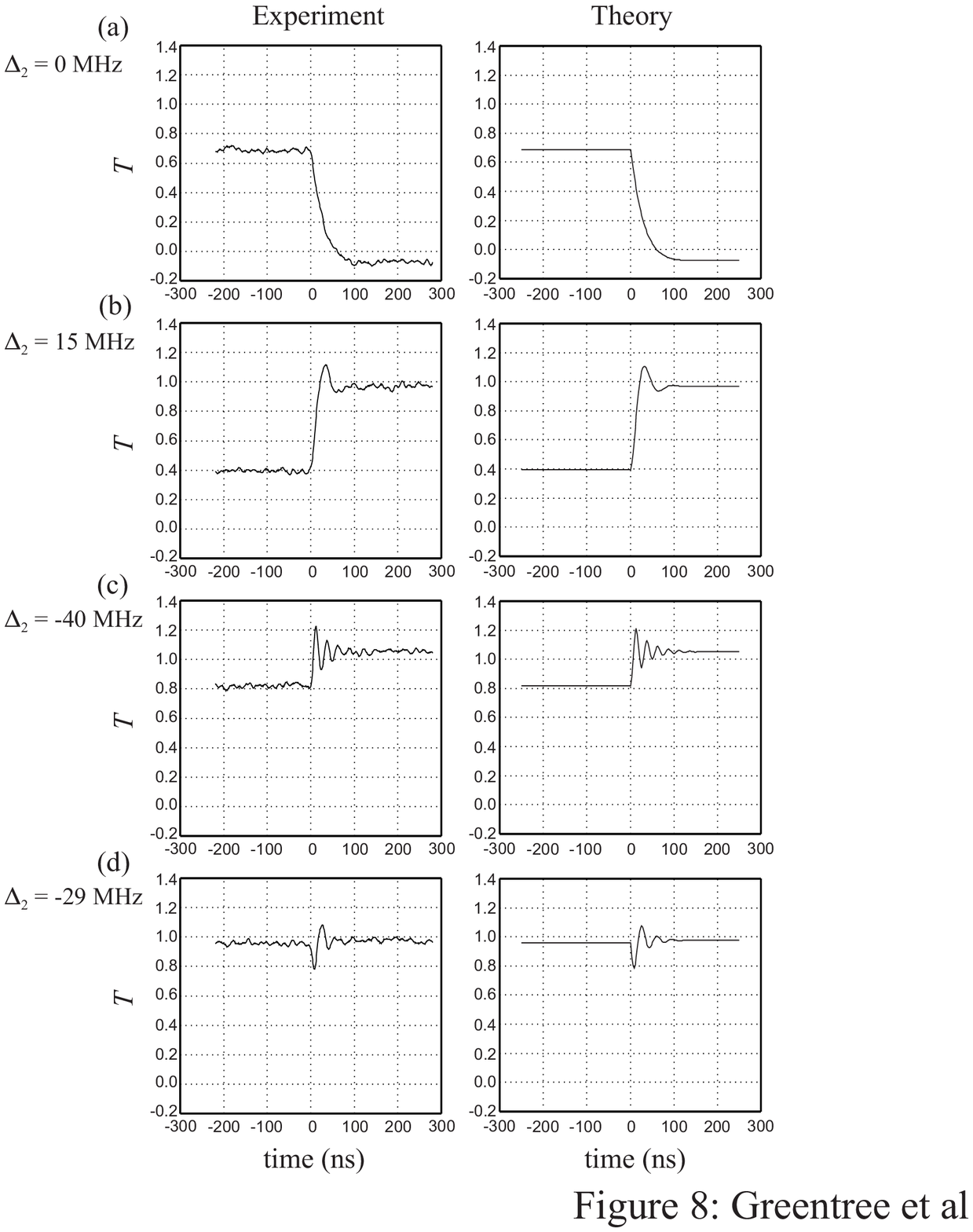}
\caption{\label{Fig8}
Turn-off transients for a strong, off-resonant coupling field: (a) 
$\Delta _{2}=0\mathop{\rm MHz}$ (bare state filling), (b) $\Delta _{2}=15%
\mathop{\rm MHz}$ (major dressed state emptying), (c) $\Delta _{2}=-40%
\mathop{\rm MHz}$ (minor dressed state emptying) and (d) $\Delta _{2}=-29%
\mathop{\rm MHz}$. \ In all cases $\Delta _{1}=-23\mathop{\rm MHz}%
$, $\Omega _{1}=46\mathop{\rm MHz}$ before turn-off and the uncoupled
absorption was found to be about $20\%$.}
\end{figure*}

\begin{figure}[tbp!]
\includegraphics[bb=146 248 450 527,width=\columnwidth,clip]{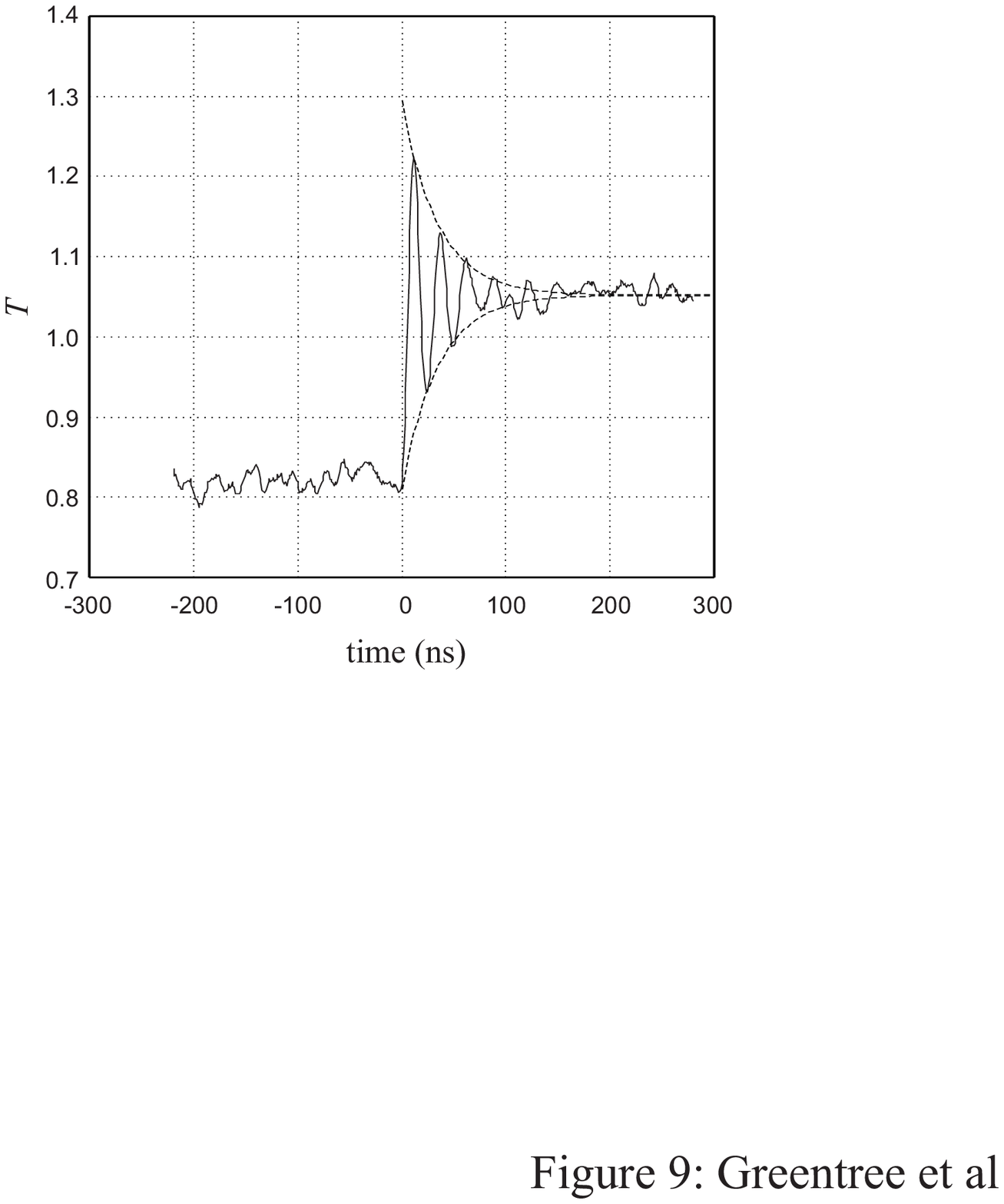}
\caption{\label{Fig9}
Close-up of experimental FIG.~\ref{Fig8}(c) showing the minor dressed
state emptying (solid curve) with added exponential decays overlayed (dashed
curves). \ The inverse lifetime of the decay was found to be $5.5\mathop{\rm MHz}%
$ which corresponds well with the measured probe spectral linewidths.}
\end{figure}

The experimental and corresponding theoretical transients for a resonant
coupling field are presented in FIGS.~\ref{Fig7}(a) to (d). \ Here $t=0$ defines
the turn-off time.\ Case (a) for a resonant probe shows the transmission
dropping from the uncoupled absorption level of $T\approx 0.6$ to $T=0$ as
the probe fills the bare state transition. \ Of more interest is case (b)
where the probe is detuned to a level corresponding to a pre turn-off
dressed state ($\Delta _{2}=-22%
\mathop{\rm MHz}%
\simeq -\Omega _{1}/2$). \ We term this dressed state {\em emptying} by
comparison with the dressed state filling reported above. \ The very clear
signal features a gain peak rising to $T=1.35$, making this another
promising region for further study of transient gain. \ We note that this is
a tunable gain region and is remote from any uncoupled absorptions. \ The
figure also shows two full periods of oscillation. \ The measured period of
oscillation was found to be $46%
\mathop{\rm ns}%
$ which is consistent with the probe detuning of $-22%
\mathop{\rm MHz}%
$ determined from the cavity off-set voltage. \ Figures~\ref{Fig7}(c) and
(d) were taken for detunings of $\Delta _{2}=17%
\mathop{\rm MHz}%
$ and $\Delta _{2}=-36%
\mathop{\rm MHz}%
$ respectively. \ These results show that cycles of gain and absorption can
be present even when there is little or no change in the steady state
transmission.

The turn-off results for a detuned coupling field are shown in FIG.~\ref{Fig8}. \
Case (a) shows the transient filling of the bare state and is similar to
FIG.~\ref{Fig7}(a). \ Cases (b) and (c) show the major and minor dressed states
emptying respectively. \ As in the case of a resonant coupling field, these
transients are characterized by ringing and gain. \ In case (b) ($\Delta
_{2}=15$MHz, corresponding to the major dressed state) these effects are
somewhat obscured by being in the wings of the uncoupled absorption. \ In
contrast, case (c) ($\Delta _{2}=-40%
\mathop{\rm MHz}%
$, corresponding to the minor dressed state) shows larger gain and
exceptionally clear ringing. \ A fit to the decay of the ringing is shown in
FIG.~\ref{Fig9}. \ As expected, this gives a decay rate of $5.5%
\mathop{\rm MHz}%
$, which corresponds to half the phenomenological decay rate measured from
the probe spectral linewidths. \ Finally, FIG.~\ref{Fig8}(d) is for a probe
detuning of $\Delta _{2}=-29%
\mathop{\rm MHz}%
$, again showing transient features between two regions of steady-state
transparency.

\section{\label{Conclusions}Conclusions}

We have carried out a theoretical and experimental study of transient
phenomena associated with the switching of the coupling field in a $\Lambda $
type EIT system. \ These studies have shown novel transient gain features
and aid the understanding of transient dressing.

We have experimentally observed interesting transient gain features in both
the turn-on and turn-off regimes. \ We have found it useful to interpret the
transient dynamics in terms of the filling and emptying of dressed states. \
The absolute peak gain values, especially the dressed state emptying, were
found to be comparable in strength to the more usual, on-resonant turn-on
case \cite{bib:OurTransient}. \ Off resonant gain peaks have an important
advantage over resonant gain peaks in that they can be tuned by appropriate
choice of coupling strengths and detunings. \ In this way they can, for
example, be removed from uncoupled absorptions or perhaps used to match
transitions of different frequencies. \ The turn-off experiments also
exhibited extremely clear transient signals and as such may be useful as a
tool to explore the properties of atomic systems.

We have obtained some new analytical results. \ In particular we give the
weak probe response after turning off the coupling field for arbitrary
detunings. \ Analysis of the resonant turn-on using the Laplace method
enabled us to identify a frequency oscillation not reported previously. \ We
analyzed this oscillation in terms of the 3D Vector Model and outlined a
parameter regime where we would expect it to be observable.

These results have applicability to the understanding of EIT for quantum
information storage and may lead the way to methods which avoid the
limitations of the condition of adiabaticity inherent in present schemes for
optical quantum information storage \cite{bib:StoppedLight,bib:FastQuantumGate}.

\begin{acknowledgements}
The authors would like to acknowledge the financial support of the EPSRC, as
well as technical support from Shahid Hanif (Imperial College, London),
Roger Bence, Fraser Robertson and Robert Seaton (The Open University, Milton
Keynes).
\end{acknowledgements}

\appendix*
\section{Further Investigations of the ``Fast'' Oscillations}

The oscillations in $%
\mathop{\rm Im}%
\left[ \rho _{bc}(t)\right] $ at the angular frequency $\Omega _{1}$ noted
in Section~\ref{Laplace} are unusual in that the Rabi frequency usually associated with
the cycling of population is approximately $\Omega _{1}/2$. The origins of
this effect in the absence of decay are easily demonstrated using the 3D
Vector Model of 3-state systems introduced recently \cite
{bib:Kasapi,bib:VDRHCH}.

The 3D Vector Model is based on the representation of the state of the atom $%
\left| \psi \left( t\right) \right\rangle =v_{a}\left| a\right\rangle
+v_{b}\left| b\right\rangle -iv_{c}\left| c\right\rangle $ by the real,
3-dimensional vector $\overrightarrow{v}(t)=(v_{a},v_{b},v_{c})^{\text{T}}$,
where here the states $\left| a\right\rangle $, $\left| b\right\rangle $ and 
$\left| c\right\rangle $ represent the bare energy eigenstates of the atom.
The components of $\overrightarrow{v}$ are given with respect to three
orthogonal axes which we shall call the $a$, $b$ and $c$ axes. These axes
represent the three energy eigenstates of the atom.

The action of the (resonant) probe and coupling fields on $\left| \psi
\right\rangle $ is represented simply as a rotation about a {\em Rabi vector 
}$\overrightarrow{\Omega }=(-\frac{1}{2}\Omega _{2},\frac{1}{2}\Omega
_{1},0)^{\text{T}}$ as follows:
\begin{equation}
\frac{d\overrightarrow{v}}{dt}=\overrightarrow{\Omega }\times 
\overrightarrow{v}.  \label{eq:Apeq1}
\end{equation}
The rotation of $\overrightarrow{v}$ about $\overrightarrow{\Omega }$ occurs
at the frequency $\Omega \equiv |\overrightarrow{\Omega }|=(\Omega
_{1}^{2}+\Omega _{2}^{2})^{1/2}/2$ which is approximately $\Omega _{1}/2$
for $\Omega _{2}\ll \Omega _{1}$. The Rabi vector lies in the $a-b$ plane at
an angle of $\theta =\tan ^{-1}(\Omega _{2}/\Omega _{1})\approx \Omega
_{2}/\Omega _{1}$ to the $b$ axis. \ The value of $\rho _{bc}$ is the
product of the amplitude of the state $\left| b\right\rangle $ and the
complex conjugate of the amplitude of state $\left| c\right\rangle $ in $%
\left| \psi \right\rangle $, which corresponds to the product $iv_{b}v_{c}$,
and so $%
\mathop{\rm Im}%
\left( \rho _{bc}\right) =v_{b}v_{c}$. \ Absorption of the probe occurs when 
$v_{b}v_{c}<0$ and gain when $v_{b}v_{c}>0$.

We note from Eq.~(\ref{eq:ImrhobcasPhis}) that, in general, $%
\mathop{\rm Im}%
\left( \rho _{bc}\right) $ has frequency components at both the $\Omega
_{1}/2$ and $\Omega _{1}$ frequencies. However, there are two important
cases where $%
\mathop{\rm Im}%
\left( \rho _{bc}\right) $ oscillates at either $\Omega _{1}$ or $\Omega
_{1}/2$.

The first case is when $\rho _{bb}^{0}=1$. Substituting this into Eq.~(\ref
{eq:ImrhobcasPhis}) gives $%
\mathop{\rm Im}%
\left( \rho _{bc}\right) $ oscillating at a frequency of $\Omega _{1}/2$. In
this case the initial state vector lies along the $\pm b$ axis, that is, $%
\overrightarrow{v}(0)=(0,\pm 1,0)^{\text{T}}$. In time the state vector $%
\overrightarrow{v}$ traces out an acute cone (of half angle $\theta $)
centered on $\pm \overrightarrow{\Omega }$ and revolves at frequency $\Omega 
$. The full solution of Eq.~(\ref{eq:Apeq1}) is given by
\begin{equation}
\overrightarrow{v}\left( t\right) =\pm \left( 
\begin{array}{c}
-\cos \theta \sin \theta \left[ \cos \left( \Omega t\right) -1\right]  \\ 
\cos ^{2}\theta +\sin ^{2}\theta \cos \left( \Omega t\right)  \\ 
-\sin \theta \sin \left( \Omega t\right) 
\end{array}
\right) . \label{eq:Apeq2}
\end{equation}
Keeping terms up to first order in $\Omega _{2}/\Omega _{1}$ gives $%
\mathop{\rm Im}%
\left[ \rho _{bc}(t)\right] \approx -(\Omega _{2}/\Omega _{1})\sin (\Omega
t)\approx -(\Omega _{2}/\Omega _{1})\sin (\Omega _{1}t/2)$ as found in
Eq.~(\ref{eq:ImrhobcasPhis}).

The second case is when $\rho _{aa}^{0}=\rho _{bb}^{0}=1/2$. In this case
Eq.~(\ref{eq:ImrhobcasPhis}) shows that $%
\mathop{\rm Im}%
\left( \rho _{bc}\right) $ oscillates only at the double frequency, i.e. $%
\Omega _{1}$. Equation~(\ref{eq:ImrhobcasPhis}) was derived for zero ground
state coherence, i.e. $\rho _{ab}^{0}=0$. This means that in the gas sample,
half the atoms are in state $\left| a\right\rangle $ and the remainder are
in $\left| b\right\rangle $. We model this by taking the average of the
values of $%
\mathop{\rm Im}%
\left( \rho _{bc}\right) $ for both collections of atoms. For the atoms
beginning in the $\left| a\right\rangle $ state, the vector is initially
aligned along the $\pm a$ axis, $\overrightarrow{v}(0)=(\pm 1,0,0)^{\text{T}}
$, and traces out a cone of half angle $\pi /2\pm \theta $. \ The solution
of Eq.~(\ref{eq:Apeq1}) for this case is given by
\begin{equation}
\overrightarrow{v}(t)=\pm \left( 
\begin{array}{c}
1+\left[ \cos \left( \Omega t\right) -1\right] \cos ^{2}\theta  \\ 
\cos \theta \sin \theta \left[ \cos \left( \Omega t\right) -1\right]  \\ 
-\cos \theta \sin \left( \Omega t\right) 
\end{array}
\right) .
\end{equation}
Keeping terms up to first order in $\Omega _{2}/\Omega _{1}$ gives $%
\mathop{\rm Im}%
\left[ \rho _{bc}(t)\right] \approx -(\Omega _{2}/\Omega _{1})\left[ \frac{1%
}{2}\sin (2\Omega t)-\sin (\Omega t)\right] $. Taking the average of this
value and the value in Eq.~(\ref{eq:Apeq2}) yields $%
\mathop{\rm Im}%
\left[ \overline{\rho _{bc}(t)}\right] \approx -(\Omega _{2}/\Omega _{1})%
\frac{1}{4}\sin (2\Omega t)\approx -(\Omega _{2}/\Omega _{1})\frac{1}{4}\sin
(\Omega _{1}t)$ as found in Eq.~(\ref{eq:ImrhobcasPhis}).

In summary, we have traced the origin of the oscillations at twice the Rabi
frequency $\Omega _{1}$ to the fact that for a pure state, $\rho _{bc}$ is
the product of the amplitude of finding the atom in state $\left|
b\right\rangle $ and the complex conjugate of the amplitude in state $\left|
c\right\rangle $. In the absence of decay, both amplitudes have terms that
oscillate at the frequency $\Omega _{1}/2$, and so their product gives rise
to a term oscillating at twice this frequency. So even though the
populations oscillate at the Rabi frequency $\Omega _{1}/2$, the absorption
and emission of light has a term which occurs at twice this frequency.

\end{document}